\documentclass[aps,showpacs,amsmath,amssymb]{revtex4}
\usepackage{graphicx}
\usepackage{dcolumn}
\usepackage{amsmath}
\usepackage{color}
\usepackage[normalem]{ulem}

\newcommand{\simgt}{\lower.5ex\hbox{$\; \buildrel > \over \sim \;$}}
\newcommand{\simlt}{\lower.5ex\hbox{$\; \buildrel < \over \sim \;$}}

\newcommand{\bfm}[1]{\mbox{${\bf #1}$}}
\newcommand{\bi}[1]{\hbox{\boldmath{$#1$}}}

\newcommand{\baredth}{\;\overline{\raise1.0pt\hbox{$'$}\hskip-6pt
\partial}\;}
\newcommand{\edth}{\;\raise1.0pt\hbox{$'$}\hskip-6pt\partial\;}

\newcommand{\gsim}{\lower.5ex\hbox{$\; \buildrel > \over \sim \;$}}
\newcommand{\lsim}{\lower.5ex\hbox{$\; \buildrel < \over \sim \;$}}

\def\aap  {A\&A}
\def\aj  {Astronomical Journal}
\def\apj  {Astrophys. J.}
\def\apjl {Astrophys. J. Lett.}
\def\apjs {Astrophys. J. Suppl.}

\def\mnras{MNRAS}
\def\nat  {Nature}
\def\prd  {Phys.~Rev.~D.}

\def\LSS  {\rm LSS}
\def\ab  {\rm ab}

\begin{document}

\title{A dipole anisotropy of galaxy distribution: Does the CMB rest-frame
exist in the local universe?}

\author{Yousuke Itoh$^1$, Kazuhiro Yahata$^2$, Masahiro Takada$^{3}$}

\affiliation{%
{}$^{1}$ Astronomical Institute, Graduate School of Science,
Tohoku University, Sendai 980-8578, Japan\\ 
{}$^2$ Department of
Physics, The University of Tokyo, Bunkyo-ku, Tokyo 113-0033, Japan \\
{}$^3$ Institute of the Physics and Mathematics of the Universe (IPMU), 
The University of Tokyo, Chiba 277-8582, Japan }
\email{yousuke@astr.tohoku.ac.jp}


\begin{abstract}
The peculiar motion of the Earth causes a dipole anisotropy modulation
in the distant galaxy distribution due to the aberration effect.
However, the amplitude and angular direction of 
the effect is not necessarily the
same as those of the cosmic microwave
background (CMB) dipole anisotropy due to the growth of cosmic
structures.
In other words exploring the aberration effect may give us a
clue to the horizon-scale physics perhaps related to the cosmic
acceleration. In this paper we develop a method to explore the dipole
angular modulation from the pixelized galaxy data on the sky properly
taking into account the covariances due to the shot noise and the
intrinsic galaxy clustering contamination as well as the partial sky
coverage. We applied the method to the galaxy catalogs constructed from
the Sloan Digital Sky Survey (SDSS) 
Data Release 6 data. 
After constructing the four galaxy 
catalogs that are different 
in the ranges
of magnitudes and photometric redshifts to study possible systematics, 
we found that the most robust sample against systematics
indicates no dipole anisotropy in the galaxy distribution. This finding is
consistent with the expectation from the concordance 
$\Lambda$-dominated cold dark matter model.  
Finally we argue that an almost full-sky galaxy survey
such as LSST may allow for a significant detection of the aberration
effect of the CMB dipole having the precision of constraining the
angular direction to $\sim 20 $ degrees in radius. 
Assuming 
a hypothetical 
LSST galaxy 
survey, we find that 
this method 
can 
 confirm or reject 
the result implied from a stacked analysis of the kinetic
Sunyaev-Zel'dovich effect of $X$-ray luminous clusters in Kashlinsky et
al. (2008,2009) if the implied cosmic bulk flow is not extended out to  
the horizon.

\end{abstract}

\maketitle

\section{Introduction}
\label{Introduction}

The amplitude of the cosmic microwave background (CMB) dipole
anisotropy is about two orders of magnitudes greater than the
anisotropies at higher multipoles that are primarily generated during
cosmic epochs until the last scattering surface of redshift 
$z\simeq 1100$. 
It is widely believed that the dipole 
anisotropy is produced
by the Doppler effect due to the relative motion between the Earth,
i.e. an observer, and the frame where the CMB looks nearly isotropic
(hereafter we call it the CMB rest-frame). 
The measured CMB dipole
amplitude tells that the relative velocity has an amplitude of $v_{\rm
CMB}\simeq 370~$km~s$^{-1}$ ($1.23\times 10^{-3}$ in the unit $c=1$)
\cite{Kogut1993,Lineweaver1996}. The peculiar velocity consists of the
five vector contributions \cite{Courteau1999}: the motion of the Earth
around the solar system barycenter
($\sim30$~km~s$^{-1}$), the motion of the solar system with respect to
the Local Standard of Rest (LSR) \cite{DehnenBinney1998}, the (hypothetical circular) motion of the LSR around the Milky Way
($\sim$220~km~s$^{-1}$ (the IAU 1985 recommended value), or $\sim$250~km~s$^{-1}$\cite{Reid2009}), the motion of the Milky Way in the
Local Group, and the motion of the Local Group with respect to the CMB
rest-frame. 
The origin of the 
fifth
component, the peculiar velocity of the Local Group,
is still uncertain and has been under discussion over
 the past two decades (e.g., \cite{Lynden-Belletal89,Straussetal92} and also see
\cite{Saleem,Erdogdu2009} and references therein). This peculiar velocity is
believed to be generated by the spatial inhomogeneities of mass (mainly
dark matter) distribution  
in
nearby large scale structures 
via gravitational instability as predicted in the cold dark matter (CDM) dominated
structure formation scenario. Therefore the peculiar velocity field is
expected to reflect properties of structure formation in the
low-redshift Universe.  
For example, the peculiar velocity of the Local Group with respect to the CMB rest-frame
is estimated to be $\sim 600$~km~s$^{-1}$, based on the result of
\cite{Courteau1999}. 
This is greater than the rms amplitude of
the peculiar velocity, $\sim 470$~km~s$^{-1}$, predicted from the linear
theory of the concordance $\Lambda$CDM model. However, due to the
difficulties in inferring the mass distribution from the observed galaxy
distribution, 
i.e.
the galaxy bias uncertainty, the origin of the peculiar
velocity  is not yet fully understood
\cite{BasilakosPlionis2006,KocevskiEbeling2006,Erdogdu2006,Watkins2009,Lavaux2010}.

Then a naive question may arise; how special is the CMB rest-frame?
One may think that cosmic large-scale structures have formed via
gravitational instability over 13.7G years in the CMB rest-frame. In
other words, the Universe may still stay in the CMB rest-frame over the
cosmic age \footnote{In General Relativity terms, the comoving
hypersurface, where the energy flux of matter comoving with the cosmic
expansion vanishes, has stayed 
the
same from the last scattering surface to 
the present-day Universe.}.
However, a possibility has been discussed that 
the CMB rest-frame may not coincide with 
the rest frame of galaxy distribution due to super-horizon scale physics
\cite{EllisBaldwin1984,Turner1991}.
In fact, the present-day Universe is in the mysterious phase, the cosmic accelerating
phase. The origin of the cosmic acceleration is one of 
the most profound problems in modern cosmology and physics, and it is sometimes discussed that the
cosmic acceleration is perhaps related to the horizon-scale physics. For
example, while dark energy is one possible explanation of the cosmic
acceleration, dark energy should have spatial perturbations inevitably
on horizon scales, if it is not a cosmological constant
(e.g. \cite{Caldwelletal98,Takada06} and references therein). Or the
primordial power spectrum, produced in the inflationary era, may have
weird behaviors on horizon scales such as the truncated power spectrum
as speculated from the low CMB quadrupole amplitudes (e.g.,
\cite{Bennettetal96,Jingetal94,Efstathiou03}). Thus exploring the
rest frame of the present-day Universe may give a clue to the
horizon-scale physics (also see \cite{Turner1991}). Furthermore,
this may give an independent test on the cosmological principle,
the isotropy and homogeneity of the Universe.

The rest frame of the present-day Universe 
may be defined by the frame where the galaxy distribution looks
isotropic on a sufficiently large scale (hereafter we will call it 
the matter rest-frame  
or the rest-frame of the local Universe). The relative
motion of the Earth to the 
matter
rest-frame causes a dipole anisotropy
modulation in the observed galaxy distribution
\cite{EllisBaldwin1984}. 
The dipole anisotropy is caused by the two
effects. First, photons from galaxies ahead of/behind the Earth are
blue-/red-shifted by the Doppler effect, causing their fluxes to be
brightened/dimmed and therefore the galaxies to be included/excluded in
the magnitude limited sample. Secondly, 
special relativity predicts
the aberration of angles 
causing 
the surface number density of
galaxies to be enhanced/suppressed in the direction ahead of/behind us,
even if the intrinsic galaxy distribution is 
perfectly homogeneous on the sky. If
the motion has a similar amplitude inferred from the CMB dipole, the
induced dipole modulation is small at a sub-percent level. Hence a
large-area survey of distant galaxies or radio sources is suited for
exploring this dipole anisotropy, because the intrinsic galaxy
clustering that has greater amplitudes at low redshifts may cause a
significant contamination. There have been many attempts made to explore
this dipole anisotropy from 
such surveys 
\cite{Baleisisetal98,Scharfetal00,BlakeWall2002}.  
In particular 
Blake and Wall \cite{BlakeWall2002} analyzed the radio source distribution
based on the NRAO VLA Sky Survey (NVSS) data \citep{NRAOVLASkySurvey},
and then claimed a possible detection of the dipole anisotropy that are
consistent with the CMB dipole in the amplitude and direction within
2$\sigma$ and 1$\sigma$ levels, respectively.

The purpose of this paper is to explore the cosmological dipole
signature from the galaxy catalogs constructed from 
 the Sloan Digital Sky Survey (SDSS) Data Release 6 (DR6) data
 \citep{York2000}.
The use of the photometric SDSS galaxy sample has several advantages:
the sky coverage is large (about 20\% of the full sky), the photometry
is well-calibrated and the photometric redshift information of each
galaxy is available.
In doing this, we develop a method to explore the dipole modulation in
a pixelized galaxy distribution properly taking into account the
covariances due to the Poisson shot noise and the intrinsic clustering
contamination as well as the partial sky coverage. In particular the
photometric redshift information of SDSS 
galaxies 
is  useful to reduce
the clustering contamination from nearby structures.  Furthermore, we
will discuss how a planned large-area galaxy survey such as the Large
Synoptic Survey Telescope (LSST) can be useful to explore the dipole
anisotropy. 

Here it may be worth mentioning advantages and shortcomings of our method compared to 
the peculiar velocity field studies 
\cite{BasilakosPlionis2006,KocevskiEbeling2006,Erdogdu2006,Erdogdu2009,Watkins2009,Lavaux2010}. 
This paper concerns with the dipole pattern in the galaxy distribution
due to the observer's motion 
with respect to the 
matter
rest-frame. 
Comparing 
this and CMB dipole 
anisotropies
can address
if the 
inferred matter rest-frame
agrees with 
the CMB rest-frame 
within the measurement errors. 
This paper
does not discuss the origin of our motion (or the motion of the Local
Group) nor 
which structures 
in the Local Group
cause our 
motion with respect to the CMB rest-frame, the so-called convergence depth. 
This 
question
is one of the main questions 
discussed in  
the recent peculiar velocity field studies 
\cite{BasilakosPlionis2006,KocevskiEbeling2006,Erdogdu2006,Erdogdu2009,Watkins2009,Lavaux2010}.

The structure of this paper is as follows. After reviewing the
aberration effect on the galaxy counts of a given magnitude limit in
Sec.~\ref{aberration}, we will develop a method to explore the
induced dipole modulation from a galaxy catalog properly taking into
account the covariance and the partial sky coverage in 
Sec.~\ref{sec:Methodology}. 
In Sec.~\ref{sec:rough_estimate}, we
make an estimate on the
detectability of the aberration effect from the galaxy distribution for
hypothetical galaxy surveys, SDSS- and LSST-type surveys, {\em assuming
} that the Earth's relative motion to 
the matter rest-frame has the 
same amplitude with the
CMB dipole. Sec.~\ref{sec:SDSS_DR6} shows the main results of this
paper. After defining the galaxy catalogs based on the magnitude range
and the photometric redshift information and then estimating the
covariance matrix of the pixelized galaxy counts on the sky, we will
show the results for an exploration of the dipole anisotropy from the
SDSS DR6 galaxy catalogs where the amplitude and angular direction of
the aberration effect are treated as free parameters. 
In Sec.~\ref{sec:FutureProspect}
 we will also   
discuss the forecast for an LSST-type survey.
Sec.~\ref{sec:Summary_Discussion}
is devoted to summary and discussion. Throughout this paper we will
employ the concordance $\Lambda$CDM model that is specified by 
$(h,\Omega_{\Lambda},\Omega_{m}h^2,\Omega_{b}h^2,\sigma_8,n_s) =
(0.73,0.762,0.127,0.0223,0.74,0.951)$ \citep{Spergel2007}.

\section{Effect of the Earth's peculiar motion on the angular number density
 fields of galaxies} \label{aberration}

Even if there exists the 
{\em matter rest-frame} 
where 
the intrinsic galaxy distribution looks perfectly isotropic,
the peculiar motion of the
Earth relative to the 
matter rest-frame induces an apparent angular
modulation in the galaxy number density field on the sky
\cite{EllisBaldwin1984}. 
There are two effects that cause this modulation.  The
first is the aberration effect, causing the observed number density
field to be increased or decreased in the angular direction forward or
backward of the Earth's motion, respectively.  The second is the Doppler
effect: depending on the shape of spectral energy density of a galaxy,
the Doppler effect causes the apparent magnitude of galaxy to be
brighter or fainter, leading the galaxy to be included into or excluded
from the magnitude limited sample. The net effect arises from the linear
sum of these two effects in a case that the peculiar velocity is much
smaller than the speed of light.

In the following, we will in more detail model the effect of the Earth's
peculiar motion on the angular number counts of galaxies 
defined for
a given limiting magnitude. In so doing, we assume for clarity that the
galaxy distribution is perfectly isotropic on the sky in the 
matter rest-frame, 
and in other words 
we ignore intrinsic anisotropies in the galaxy
distribution arising from large-scale structure formation.  We shall
come back to the intrinsic anisotropy contamination later.

Special Relativity predicts that, due to aberration of angle, the
relative velocity of an observer to the 
matter rest-frame causes the
angular position of a galaxy observed on the sky to be displaced:
\begin{eqnarray}
\tan\phi_{\rm obs} &=& 
\frac{\sqrt{1-\beta^2} \sin\phi
}{\cos\phi +\beta}\approx  \frac{\sin\phi
}{\cos\phi +\beta}, 
\label{eqn:tant}
\end{eqnarray}
where $\phi_{\rm obs}$ denotes the angle between the direction of the
galaxy seen by the observer and the direction of the observer's velocity
$\bfm{v}$, $\phi$ is the corresponding angle in the 
matter rest-frame,
and $\beta\equiv |\bfm{v}|/c$. In the second equality on the r.h.s. we
assumed $\beta\ll 1$, 
and
ignored the term of $O(\beta^2)$.  In the following we will similarly
ignore the contribution of $O(\beta^2)$ for simplicity.  The
conservation of the number of galaxies tells that
the observer sees an
angular modulation in the galaxy distribution as a function of the
angular direction $\bi{\theta}$ on the celestial sphere:
\begin{equation}
n(\bi{\theta})=\frac{d \phi}{d \phi_{\rm obs}}
\approx \bar{n}\left(1+2\beta \cos\alpha \right),
\label{eqn:n_ab}
\end{equation}
where ${\bi{\theta}}=
(\cos\theta\cos\varphi,\cos\theta\sin\varphi,\sin\theta)$ for the
spherical coordinates 
(we have employed the continuous field limit,
which is a good approximation for cases of interest)
and $\bar{n}$ is the intrinsic number counts per unit
steradian. The angle $\alpha$ is defined by $\cos\alpha\equiv
\bi{\theta}\cdot\hat{\bfm{v}}$, where $\hat{\bfm{v}}$ denotes the
angular direction of the Earth's peculiar velocity $\bfm{v}$ on the celestial
sphere and is fully specified by two parameters. 

Next let us consider the Doppler effect on the galaxy counting.  Besides
redshift due to the cosmic expansion, the peculiar velocity of an
observer relative to the comoving rest frame of galaxy distribution
causes a photon emitted from a galaxy to be redshifted or blueshifted
depending on 
the angular direction of the galaxy relative to the peculiar velocity 
direction. 
The observed frequency $\nu_{\rm obs}$ is related
to the rest frame frequency $\nu_{\rm rest}$ via
\begin{equation}
\eta \equiv \frac{\nu_{\rm obs}}{\nu_{\rm rest}}=
 \frac{1+\beta \cos\phi}{(1-\beta^2)^{1/2}}\approx 1+\beta\cos\phi,
\label{eqn:eta}
\end{equation}
where the angle $\phi$ is defined in the same way as in
Eq.~(\ref{eqn:tant}). If we assume that an intrinsic flux density of a
galaxy [in the units of erg~s${}^{-1}$~cm$^{-2}$~Hz$^{-1}$] is simply
given by a power law as $S_{\rm rest}(\nu)\propto \nu^{p}$, the
conservation of photon number tells that the observed flux density
$S_{\rm obs}(\nu)$ is expressed as
\begin{equation}
S_{\rm obs}(\nu_{\rm obs})
=S_{\rm rest}(\nu_{\rm rest})\frac{\nu_{\rm obs}}{\nu_{\rm rest}}
\frac{d\nu_{\rm rest}}{d\nu_{\rm obs}}\frac{d t_{\rm rest}}{d t_{\rm obs}}
=\eta^{1-p}S_{\rm rest}(\nu_{\rm obs}).
\end{equation}
The apparent magnitude of the galaxy in a given filter is obtained by
integrating the flux density over a range of transmission frequencies of
the filter, and therefore the observed magnitude, $m_{\rm obs}$, is
related to the rest frame magnitude, $m_{\rm rest}$, via
\begin{eqnarray}
m_{\rm obs}
&=&m_{\rm rest}-2.5(1-p)\log_{10}\eta.
\label{eqn:m_obs}
\end{eqnarray}
If we assume that the intrinsic number counts of galaxies, where galaxies
brighter than a given limiting magnitude $m_{\rm lim}$ are included in
the sample, is simply given by
\begin{equation}
\bar{n}(m<m_{\rm lim})\propto 10^{x m_{\rm lim}},
\end{equation}
with $x$ 
being
a numerical coefficient of order unity \cite{Fukugita2004},  
the observed number counts are found from Eqs.~(\ref{eqn:eta}) 
and 
(\ref{eqn:m_obs}) 
to be 
\begin{eqnarray}
n(\bi{\theta}; m<m_{\rm lim})&=&\bar{n}(m<m_{\rm rest, lim}=m_{\rm lim}+2.5(1-p)\log_{10}\eta)
\nonumber\\
&=&\eta^{2.5x(1-p)}\bar{n}(m<m_{\rm lim})
\nonumber\\
&\approx& \left[1+2.5x(1-p)\beta\cos\alpha\right]\bar{n}(m<m_{\rm lim}), 
\label{eqn:n_dop}
\end{eqnarray}
where the angle $\alpha $ is defined in the same way as in
Eq.~(\ref{eqn:n_ab}). 

Hence, taking into account both the aberration effect (\ref{eqn:n_ab})
and the Doppler effect (\ref{eqn:n_dop}) up to the first order of  
$\beta$, the observed angular number density field of galaxies for the
limiting magnitude $m_{\rm lim}$ is expressed as
\begin{equation}
n(\bi{\theta}; m<m_{\rm lim})=\bar{n}(m<m_{\rm lim})
\left[1+2\tilde{\beta}\cos\alpha\right],
\label{eqn:del-rel}
\end{equation}
with the modified $\beta$ parameter defined as 
\footnote{In fact, we will consider flux limited as well as redshift limited
samples in this paper (see Table \ref{tab:samp}). 
Because redshifts of galaxies ahead of our motion 
have systematically smaller than those behind us, there arises 
an additional dipolar variation in the number density. Up to 
$O(\beta)$ and assuming we study galaxies in the redshift range $z_1 \le
z \le z_2$, 
we can take account of this effect by writing
$$
\tilde{\beta} =
\left(1+1.25x(1-p) + F(z_1,z_2)\right)\beta. 
$$
instead of Eq. (\ref{eqn:tilde_beta}) where 
$$
2 F(z_1,z_2) = \left[(1+z_1)n_g(z_1) - (1+z_2)n_g(z_2)
+ \int^{z_2}_{z_1}n_g(z)dz 
\right], 
$$
\noindent 
where $n_g(z)$ is the galaxy redshift distribution in the matter rest 
frame. Assuming Eq. (\ref{eq:nofz}) for $n_g(z)$ with 
$z_0 = 0.1$ for SDSS or $z_0 = 0.4$ for LSST, one can show that 
$F(0.1,0.4) = 0.33$ (SDSS shallow sample), 
$F(0.1,0.9) = 1.4$ (SDSS deep sample), or 
$F(1,\infty) = 0.91$ (LSST).
Hence $F(z_1,z_2)$ is positive and $O(1)$ in the case studies 
in this paper. 
}
\begin{equation}
\tilde{\beta}=\left[1+1.25x(1-p)\right]\beta. 
\label{eqn:tilde_beta}
\end{equation}
Note again that the dependence of $n(\bi{\theta})$ on angular
direction $\bi{\theta}$  comes through the relation
$\cos\alpha\equiv {\bi{\theta}}\cdot\hat{\bfm{v}}$.
Eq.~(\ref{eqn:del-rel}) shows that the Earth's motion relative to the
matter rest-frame induces 
a characteristic dipole pattern in the angular
number density field of galaxies.  The dipole pattern is fully specified
by 3 parameters: $\tilde{\beta}$ and
2 parameters for $\hat{\bfm{v}}$.
The average of Eq.~(\ref{eqn:del-rel}) over the whole sky indeed
satisfies conservation of the total number of galaxies:
\begin{eqnarray}
\frac{1}{4\pi}\oint_{4\pi}\!d\Omega_{\bi{\theta}}~ n({\bi{\theta}})
= 
\bar n. 
\label{eqn:conserv}
\end{eqnarray}

If the matter rest-frame is 
the
same as the CMB rest-frame with respect to
which the Earth is moving with the velocity 
$\beta\approx 
(1.231\pm 0.008)
\times 10^{-3}$ in the direction $(l,b) =
(264.\!\!^{\circ}31\pm 0.\!\!^{\circ}04 \pm
0.\!\!^{\circ}16,48^{\circ}\!\!.05\pm 0^{\circ}\!\!.02 \pm
0^{\circ}\!\!.09)$ 
in the Galactic coordinates 
\citep{Kogut1993,Lineweaver1996}, the dipole amplitude is 
$O(\beta)\sim O(10^{-3})$ as the prefactor in
Eq.~(\ref{eqn:tilde_beta}) in front of $\beta$ is of the order of unity
for the number counts of galaxies in optical 
passbands
\cite{Fukugita2004,BruzualCharlot2003,Stabenau2008}. 
To
detect the dipole pattern in the galaxy distribution, a galaxy survey
with full sky coverage is ideally needed, while a partial-sky survey
such as SDSS makes it less straightforward to explore the dipole
pattern, because the galaxy distribution displays a smaller angular
modulation over the sky region observed,
as will be discussed below.

\section{Methodology: a $\chi^2$ test for detecting the aberration effect}
\label{sec:Methodology}

In reality, a galaxy distribution actually seen is quite far from 
homogeneous,
rather displaying rich, hierarchical
structures on various distance scales -- cosmic large-scale
structures. To measure the aberration effect due to the Earth's peculiar
velocity, we need to discriminate the effect from the 
inhomogeneities 
due to 
large-scale structures. In addition we have to
take into account observational effects such as the survey geometry and
the shot noise contamination due to a finite number of galaxies. In this
section we develop a methodology 
for measuring 
the aberration effect
from a wide-field galaxy survey, which will be applied to the SDSS data
in 
subsequent sections.

For an actual galaxy survey, the density perturbation field has to be
estimated from the discrete distribution of galaxies.  Since the dipole
modulation of interest appears over an angular scale of $\pi$ radian and
we are not interested in small angular scales, it is convenient to
consider a pixelized map of galaxy distribution, where the pixel size
greater than degree scales would be sufficient for our purpose.
We define the number density of galaxies in the $i$-th pixel as
\begin{equation}
n_{\rm obs}(\bi{\theta}_i)=
\sum_{j=1}^{N_{\rm gal, tot}} 
\int_{\Omega_{\rm S}}\!\!d^2\Omega_{\bi{\theta}}~ 
W_{(i)}(\bi{\theta}_i-\bi{\theta})
\delta^2_D(\bi{\theta}-\bi{\theta}_{g,j}),
\label{eq:pixelizedNumber}
\end{equation}
where $\bi{\theta}_i$ denotes the angular position of the $i$-th pixel,
given by $\bi{\theta}_i=
(\sin\theta_i\cos\varphi_i,\sin\theta_i\sin\varphi_i,\cos\theta_i)$ in
the celestial polar coordinates, $\delta_D^2(\bi{\theta})$ is the Delta
function, $\bi{\theta}_{g,j}$ denotes an angular position of the $j$-th
galaxy, and the summation $\sum_j$ runs over all the galaxies used in
the analysis, $j=1,2,\dots, N_{\rm gal, tot}$ ($N_{\rm gal, tot}$ is the total
number). Note that the integration $\int_{\Omega_{\rm
S}}\!\!d^2\Omega_{\bi{\theta}}$ is confined to the survey region 
with 
area $\Omega_{\rm S}$.
The window function $W_{(i)}(\bi{\theta})$ defines the pixel shape
normalized as $\int\!\!d^2\bi{\theta}~ W_{(i)}(\bi{\theta})=1$, where
the subscript $(i)$ is used for notational convenience to explicitly
imply that the window function may change by pixel-to-pixel taking into
account possible variations due to the pixel geometry,  masking and
so on.  For example, a top-hat type window function is given by
$W_{(i)}(\bi{\theta})=1/(\pi \theta_{{\rm pix}(i)}^2)$ if
$|\bi{\theta}|<\theta_{{\rm pix}(i)}$, otherwise zero.

For our purpose we need to deal with the density perturbation field
defined as  
\begin{eqnarray}
\delta_{\rm obs}(\bi{\theta}_i)=\frac{n_{\rm obs}(\bi{\theta}_i)}
{N_{\rm gal, tot}/\Omega_{\rm S}}-1, 
\label{eqn:delta_obs}
\end{eqnarray}
Note that the 
$N_{\rm gal, tot}/\Omega_{\rm S}$
is an estimate on the average number density for a
given survey.

Accordingly we need to modify a modeling of the aberration effect such
that the model prediction can be compared with the pixelized galaxy
distribution above: the number density in the $i$-th pixel can be simply
modeled from Eq.~(\ref{eqn:del-rel}) as
\begin{equation}
n_{\rm model}(\bi{\theta}_i)=\bar{n}\left[
1+2\tilde{\beta}\cos\alpha_i
\right], 
\label{eqn:n_mod1}
\end{equation}
where $\cos \alpha_i\equiv \bi{\theta}_i\cdot\hat{\bfm{v}}$. 
However, when a survey has a partial sky coverage, the conservation law
(\ref{eqn:conserv}) no longer holds. 
In other words, 
\begin{eqnarray}
\int_{\Omega_S}\frac{d\Omega_{\bi \theta}}{\Omega_S}
n({\bi \theta}) &=&  
\bar n \left( 1 + 2\tilde \beta 
\int_{\Omega_S}\frac{d\Omega_{\bi \theta}}{\Omega_S}
\cos\hat {\mathbf v}\cdot {\bi \theta}
\right) \ne \bar n, 
\end{eqnarray}
where $\Omega_S$ is the survey area, 
and the integration range is confined to the survey region.
Hence, the average number
density $\bar{n}$ in Eq.~(\ref{eqn:n_mod1}) needs to be estimated taking
into account the partial sky coverage:
the model
density perturbation field, just like the derivation in
Eq.~(\ref{eqn:delta_obs}), is modified as
\begin{eqnarray}
\delta_{\rm model}(\bi{\theta}_i)&\equiv&
\frac{1+2\tilde{\beta}\cos\alpha_i}{ (1/\Omega_{\rm S})\sum_{j=1}^{N_{\rm
pix}}\Omega_{{\rm pix}(j)}(1+2\tilde{\beta}\cos\alpha_j) }-1\nonumber \\ 
&\simeq&
2\tilde{\beta} \left[\cos\alpha_i-\frac{1}{\Omega_{\rm
S}}\sum_{j=1}^{N_{\rm pix}}\Omega_{{\rm pix}(j)} \cos\alpha_j \right],
\label{eqn:delta_model}
\end{eqnarray}
where the summation runs over all the pixels used ($N_{\rm pix}$ is the
total number).  Note that the denominator in the first line gives an
estimate on the average number density for a case of the partial sky
coverage, and the averaging of the aberration effect is weighted by the
pixel area, $\Omega_{{\rm pix}(j)}$.

In this paper, to measure the aberration effect for a given galaxy
survey, we employ a simplest statistic $\chi^2$.  From
Eqs.~(\ref{eqn:delta_obs}) and (\ref{eqn:delta_model}), the
$\chi^2$ statistic can be given by
\begin{eqnarray}
\chi^2(\hat{\bfm{v}}, \tilde \beta)&\equiv& \sum_{i,j=1}^{N_{\rm pix}}
\left[
\delta_{\rm obs}(\bi{\theta}_i)-\delta_{\rm model}(\bi{\theta}_i; \hat{\bfm{v}},
\tilde \beta)
\right]
\left[\bfm{C}\right]^{-1}_{ij}
\left[
\delta_{\rm obs}(\bi{\theta}_j)-\delta_{\rm model}(\bi{\theta}_j; \hat{\bfm{v}},
\tilde \beta)
\right],
\label{eq:chi2definition}
\end{eqnarray}
where $[\bfm{C}]^{-1}$ is the inverse of the covariance matrix as
explained below.  The best-fit model parameters for $\hat{\bfm{v}}$ and
$\tilde{\beta}$, 3 parameters in total, can be estimated by minimizing
the $\chi^2$ value with varying the model parameters freely, given the
pixelized galaxy distribution.

The statistical uncertainty in measuring the aberration effect is
quantified by the covariance matrix $\bfm{C}$ in
Eq.~(\ref{eq:chi2definition}).  Following \cite{Peebles80} (see Sec.~31;
also see \cite{TakadaBridle07}), the covariance matrix $\bfm{C}_{ij}$ is
found to be given by
\begin{equation}
\bfm{C}_{ij}=\frac{\delta^K_{ij}}{N_{g(i)}}+w_g(\theta_{ij}),
\label{eqn:cov}
\end{equation}
where 
$\theta_{ij}\equiv
\cos^{-1}(\bi{\theta}_i\cdot\bi{\theta_j})$, 
$\delta^K_{ij}$
is the Kronecker delta function, $N_{g(i)}$ is the number of galaxies
contained in the $i$-th pixel, and $w_g(\theta)$ is the angular
two-point correlation function of the pixelized galaxy distribution. The
first term gives the shot noise contamination due to a finite number of
galaxies, while the second term quantifies the sampling variance
originating from 
the intrinsic galaxy clustering in 
large-scale structure. Note that the second term is non-vanishing 
when
$i\neq j$, which describes 
how the errors between different pixels are
correlated with each other.  
The covariance matrix $\bfm{C}$ has a dimension of $N_{\rm
pix}\times N_{\rm pix}$.

The two-point correlation function $w_g(\theta)$ in Eq.~(\ref{eqn:cov})
can be expressed in terms of the angular power spectrum of galaxy
distribution, $C_g(l)$, as
\begin{eqnarray}
w_g(\theta_{ij}) &=& 
\sum^{\infty}_{l=1}\frac{2l+1}{4\pi}C_g(l) \tilde{W}_{(i)}(l\Theta_{{\rm
pix}(i)})
\tilde{W}^\ast_{(j)}(l\Theta_{{\rm
pix}(j)})
P_l(\cos\theta_{ij}),
\label{eq:angularPowerSpectrum}
\end{eqnarray} 
where 
$\tilde{W}_{(i)}(x)$ is the Fourier transform of the $i$-th pixel's
window function $W_{(i)}(x)$, and $P_l(x)$ is the $l$-th order Legendre
polynomial function. The symbol $^*$ denotes 
the complex conjugate.
Assuming a linear bias 
model between the matter and galaxy distributions, which is
a good approximation on angular scales of interest, the angular power
spectrum $C_g(l)$ can be expressed in terms of the underlying linear
power spectrum of mass perturbations as
\begin{eqnarray}
C_g(l) &=& \frac{2}{\pi}b^2_g
\int^{\infty}_0
P^L_m(k)|I_l(k)|^2
k^2 dk,
\label{eq:angularpowerspectrum} 
\end{eqnarray} 
with
\begin{eqnarray}
I_l(k) &=& \int^{\infty}_0 D(z)n_g(z)j_l(kr(z))dz, 
\label{eq:angularpowerspectrum_kernel} 
\end{eqnarray} 
where $b_g$ is the linear bias parameter of galaxies, $r$ is the
comoving angular-diameter distance, $D(z)$ is the linear growth rate
normalized as $D(z=0)=1$ today,  $n_g(z)$ is the redshift distribution of
galaxies normalized as $ \int^{\infty}_0 n(z) dz = 1, $ $j_l(x)$ is the
$l$-th order spherical Bessel function,
and $P^L_m(k)$ is the linear mass power spectrum today.  The CDM based
linear perturbation theory provides secure predictions for $P^L_m(k)$ as
a function of cosmological parameters \citep[e.g.,
see][]{Dodelson2003}. Once the CDM model and the galaxy bias are assumed
and the galaxy redshift distribution is estimated or known, one can make
a secure estimate of the sample variance contribution to the
covariance. We will use the FFTLog code developed in
\cite{Hamilton2000} to compute the $z$-integration in
Eq.~(\ref{eq:angularpowerspectrum_kernel}) (also see \cite{Tegmark2002}).  An
alternative method is to use the angular correlation function measured
from the survey 
itself, 
however, it is generally difficult to obtain an 
accurate measurement of $w_g(\theta)$ on large angular scales 
we are
interested in.

\section{A rough estimate on the signal-to-noise for a galaxy survey}
\label{sec:rough_estimate}

\begin{table*}
\begin{center}
\begin{tabular}{l|ccccccc}
\hline\hline
Survey & Area ($\Omega_{\rm S}$) &
$N_{\rm gal, tot}/\Omega_{\rm S}$ &
$z_{\rm m}$ &
$A_{\rm ab}$ &
$A_{\rm LSS}$ &
$A_{\rm P}$ &
$S/N$\\
&
(deg$^2$) &
(deg$^{-2}$)&
&
&
&
&\\
\hline
SDSS-like ($f_{\rm sky}\approx 0.19$)&
7838& 
487&
$\sim 0.3$&
$5.3\times 10^{-4}$&
$1.3\times 10^{-3}$&
$2.8\times 10^{-4}$&
0.40 \\
low-$z$ full-sky ($f_{\rm sky}=1$)&
41200& 
487&
$\sim 0.3$&
$2.4\times 10^{-3}$&
$2.8\times 10^{-3}$&
$6.6\times 10^{-4}$&
0.84 \\
LSST-like ($f_{\rm sky}\simeq 0.46$)&
19000& 
$9.7\times 10^4$&
$1.2$&
$1.2\times 10^{-3}$&
$2.1\times 10^{-4}$&
$3.2\times 10^{-5}$&
5.5\\
\hline\hline
\end{tabular}
\end{center}
\caption{
A rough estimates of the expected signal to noise ratios for three survey
 configurations.  The ``SDSS-like'' survey assumes 
the survey geometry, 
 galaxy redshift distribution and  galaxy number density 
of  
the SDSS DR6 for survey parameters. 
The ``low-z full-sky'' survey assumes the full sky 
coverage, but 
other parameters are the same as  
the SDSS parameters.
The ``LSST-like'' survey assumes the configuration of the future LSST
 which will be explained in Sec. \ref{sec:FutureProspect}. The notations are 
 as follows. 
$f_{\rm sky} = \Omega_{\rm S}/(4\pi)$: the sky coverage of the survey. 
$N_{\rm gal,tot}/\Omega_{\rm S}$: the average number 
density of galaxies per unit square degrees.
$z_{\rm m}$: the mean redshift of the galaxy
 distribution.
$A_{\rm ab}$: the dipole amplitude coefficient of the
 aberration effect assuming the 
CMB dipole amplitude,
$\tilde \beta =  \beta\approx 1.231\times 10^{-3}$, where
the
 survey geometry is also taken into account. 
$A_{\rm LSS}$: the dipole amplitude coefficient expected
 from angular galaxy clustering in large-scale 
structures.
$A_{\rm P}$: the dipole amplitude coefficient expected
 from the Poisson noise due to a finite number density of galaxies.
$S/N$: the estimated signal-to-noise ($S/N$) ratio for
 measuring the aberration effect. 
\label{tab:sn}}
\end{table*}
Before going to the results, it would be useful to have a rough estimate
on the signal-to-noise ratio for measuring the aberration effect from a
given galaxy survey.
As discussed in Sec.~\ref{sec:Methodology}, the measurement is
contaminated by the shot noise of discrete galaxy distribution and the
galaxy clustering contribution in large-scale structures. Here we simply
compare the expected dipole amplitude due to the aberration effect with
those of the contaminating effects, taking into account 
the survey geometry
(see \cite{Baleisis1998} for the details).
A detailed derivation of the equations used in this section 
is also given in the appendix \ref{sec:appendix:Dipole_Amplitudes}.

As partially discussed in Sec.~\ref{sec:Methodology}, an incomplete sky
coverage dilutes the dipole modulation in the galaxy distribution.
For the case of a partial survey coverage, 
the dipole amplitude of the aberration effect is estimated
as 
\begin{equation}
A_{\rm ab}
\equiv
2\tilde \beta \sqrt{\frac{S_{11}}{3}}.
\label{eqn:a_ab}
\end{equation}
The quantity $S_{11}$ takes into account the
dilution effect due to the survey geometry
and is found to be given by
\begin{equation}
S_{ll'}\equiv \sum_{m=-l}^l\sum_{m'=-l'}^{l'}|W_{ll'}^{mm'}|^2,
\end{equation}
with the window function $W_{ll'}^{mm'}$ being defined 
in \citep{Peebles1973} as  
\begin{equation}
 W_{ll'}^{mm'}\equiv \oint\!\!d\Omega_{\bi{\theta}}~ 
Y^m_l(\bi{\theta}) Y^{m'\ast}_{l'}(\bi{\theta}) M(\bi{\theta}), 
\end{equation}
where $Y_{l}^m$ is the spherical harmonic function and $M(\bi{\theta})$
is the mask function to define the survey region: $M(\bi{\theta})=1$ if
the angular position $\bi{\theta}$ is inside the survey region,
otherwise zero.

The dipole component arising from the clustering distribution of
galaxies can be characterized by the angular power spectrum, weighted
with 
the survey
window function. Following the method developed
in \cite{Baleisis1998}, the dipole amplitude can be estimated as
\begin{equation}
A_{\rm LSS}=\sqrt{\frac{3}{4\pi}}\sqrt{\sum_{l'=1}^\infty
S_{1l'}C_g(l')}, 
\label{eqn:a_lss}
\end{equation}
where the prefactor $\sqrt{3/4\pi }$ is from the definition of the
first-order spherical harmonic function,
$Y_1^0=\sqrt{3/4\pi}\cos\theta$. The partial sky coverage causes the
power spectra of all the $l'$-th orders, $C_g(l')$,  
to contribute to the dipole amplitude.

The dipole amplitude arising from the shot noise due to the discrete
galaxy distribution is similarly estimated as
\begin{equation}
A_{P}=\frac{3}{\sqrt{4\pi}} \sqrt{
\frac{\Omega_{\rm S}}{4\pi \bar n_{\rm g}}}.
\label{eqn:a_p}
\end{equation}
Note that $1/\bar n_{\rm g}$ is the inverse of the
average angular number density of galaxies per steradian.

Hence, from Eqs.(\ref{eqn:a_ab}), (\ref{eqn:a_lss}) and
(\ref{eqn:a_p}), 
the signal-to-noise ratio for measuring the aberration effect may
be estimated as
\begin{equation}
\frac{S}{N}=\frac{A_{\rm ab}}{\sqrt{A_{\rm LSS}^2+A_{\rm P}^2}}. 
\label{eqn:sn}
\end{equation}
The $S/N$ simply assesses the dipole amplitude of the aberration effect
relative to those of the contaminating effects.

Table~\ref{tab:sn} gives the estimates on $S/N$ for three types of
galaxy surveys assuming the CMB dipole amplitude 
for the aberration effect,
 assuming $\tilde \beta =  \beta=1.231\times 10^{-3}$.  
First we consider a `SDSS-like survey' that mimics the
SDSS galaxy distribution analyzed in this paper, including the survey
geometry, the total number of galaxies and the redshift distribution.
In this case $S/N\simeq 0.4$, smaller than unity, meaning that the
aberration effect is difficult to measure. Comparing the dipole
amplitudes, $A_{\rm ab}$, $A_{\rm LSS}$, and $A_{\rm p}$ clarifies
that $A_{\rm LSS}$ gives a significant 
contamination to an extraction of the aberration effect.
To study the impact of the partial sky coverage, 
the row labeled as `full-sky' shows the
results when considering the full-sky coverage but keeping other survey
parameters to be 
the
same as those for the SDSS-like survey. The aberration
signal becomes more significant, however, the clustering dipole
amplitude contamination is still significant, 
resulting only in
a slight improvement as $S/N\simeq 0.8$.  On the other hand, the LSST-like
survey that probes the galaxy distribution up to much higher 
redshifts and with a wide
sky coverage $f_{\rm sky}\approx 0.46$ is found to allow for a significant
detection of the aberration effect, $S/N\simeq 5.5$, more than
$5\sigma$.  Therefore these results show that it is more important to
probe the galaxy distribution at {\em higher} redshifts for measuring
the aberration effect
mainly because of the following two reasons. 
First, the galaxy distribution is less evolving and more in
the linear regime at higher redshifts, therefore the clustering dipole 
less contaminates to the
dipole signal. Second, while the cosmic structures at smaller distance
scales are more rapidly evolving towards the nonlinear regime in the CDM
structure formation scenario, the small-scale structures at {\em higher}
redshifts are viewed by an observer with smaller angles, less
contributing to the dipole amplitude appearing at large angular scales. 

\section{Application to SDSS DR6 photometric galaxies sample}
\label{sec:SDSS_DR6}

The SDSS 
6th Data Release (DR6) \cite{York2000,Adelman2007}
 covers about 8000 deg$^2$ of sky area and contains over 
200 million of objects with photometry in five pass bands: 
$u,g,r,i$, and $z$ 
\citep{Fukugita1996,Gunn1998}.
The SDSS galaxy sample would be a most suitable data set to explore the
aberration effect because of
the well-calibrated, 
homogeneous
photometric and astrometric properties.

\subsection{Constructing Galaxy Samples from the SDSS Photometric Galaxy Catalog}
\label{sec:SDSS_DR6_Data_Reduction}

We construct the galaxy sample from the SDSS photometric catalog
following \cite{Yahata2007}.
There are several uncertainties in the
photometric calibration that may cause artificial angular modulations in
the galaxy distribution, such as an imperfect correction of Galactic dust
extinction and 
the contamination of stars to the sample caused by an
imperfect start-galaxy separation. We will also use photometric redshift
(hereafter, simply photo-$z$)
information of each galaxy  to define our galaxy sample.  We will below
briefly describe the definition of our galaxy sample.

\subsubsection{Dust extinction correction}

An inaccurate correction of Galactic dust extinction may cause artificial angular
modulations in the galaxy distribution that is defined for a magnitude-limited
sample.
In the SDSS database, each object has information on 
not only its photometric properties but also the dust extinction
estimated based on the Galactic dust extinction
map in \cite{Schlegel1998} (hereafter SFD), $A_{x,SFD}$, ($x=u,g,r,i$,
and $z$). 
As carefully investigated in \cite{Yahata2007}, there may remain a small
systematic bias in the extinction correction, especially in the field
with $A_{r,SFD} < 0.1$.  However, the possible systematics has not been
yet resolved, 
so we adopt the magnitudes where
dust extinction was
corrected for based on the SFD map. 

\subsubsection{Star-galaxy separation}
\label{subsubsec:stargalaxyseparation}

A secure galaxy-star separation is also important, because star
contamination to the galaxy sample likely causes angular modulations in
the galaxy number counts towards the Galactic plane. 
We constructed a photometric galaxy sample taking
the following three steps: 
\begin{enumerate}
\item False objects were discarded using photometric processing flags.
Namely, we removed 
the photometric objects that
 have saturated fluxes, were observed during bad sky
      conditions or are identified as fast-moving objects.
\item Masked regions were excluded,
where the masked regions are defined from regions labeled 
as``BLEEDING'', ``BRIGHT
      STAR'', ``TRAIL'', or ``HOLE''.
\item The magnitude range to define the galaxy sample was employed to
ensure a reliable  star-galaxy separation.
\end{enumerate} 
The details on Step 1 and 2 can be found in \cite{Yahata2007}.
In Step 3 we employed the two $i$-band magnitude ranges:
$19.1 \le m_i \le 19.6$ and $19.6 \le m_i \le 20.1$, respectively.

\subsubsection{Reducing a clustering dipole}
\label{subsubsec:reducing_a_clustering_dipole}

Nearby non-linear structures are viewed by an observer on relatively
larger angular scales after projection.  This causes a contamination to
our seeking of the aberration effect because the galaxy clustering on
small scales contaminates to the dipole amplitude for a survey with
partial sky coverage. 
For example, the apparent
galaxy distribution in 
the nearby large scale structures, 
which extends up to $\sim 60 -
200$ Mpc$h^{-1}$ in radial distance, is away from the CMB dipole direction
only by $\sim 10 - 20$ degrees on the sky
\citep[e.g.][]{BasilakosPlionis2006,KocevskiEbeling2006,Erdogdu2006}.
In addition, from the SDSS galaxy catalog itself, 
the apparent galaxy concentration
at $z\sim 0.08$, the so-called ``Sloan Great Wall'', has
been found \citep{Gott2005}, and is  
away from the CMB dipole
direction only by $\sim 10$ degrees.
Thus including such nearby non-linear structures in the galaxy sample
may apparently enhance large-angle clustering amplitudes in the galaxy
distribution toward  the local structures, 
which in turn prevents us from detecting the
aberration effect due to the Earth's peculiar motion relative to the
rest-frame of the SDSS galaxy distribution at typical redshifts $z\sim
0.3$. Hence, to maximize a chance to detect the aberration effect on the
SDSS galaxy sample, it is desirable to remove such 
nonlinear structures at low redshifts as
much as possible. 

\begin{center}
\begin{figure}
\centering\includegraphics[width=9cm]{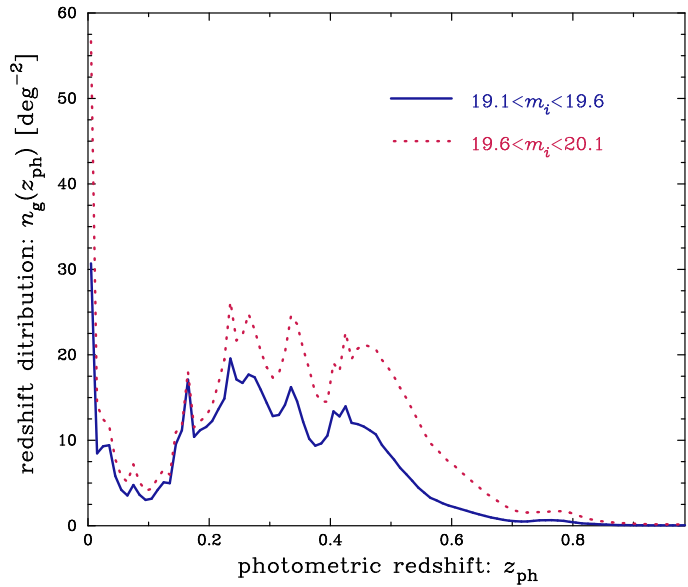}
\caption{
The photometric redshift distribution ($0 \le z \le 1$) of 
galaxies used in this analysis, from the SDSS DR6 data. 
The solid curve shows the distribution for galaxies with $i$-band
 magnitude in the range $19.1\le m_i\le 19.6$, while the dotted curve
 for $19.6\le m_i\le 20.1$.}
\label{fig:photoz}
\end{figure} 
\end{center}

For the reasons mentioned above, 
we use photometric redshift information to define a 
secure galaxy
sample.
Fig.~\ref{fig:photoz} shows the photometric redshift
distribution of  SDSS galaxies per unit steradian as a
function of $i$-band magnitudes. One can see 
that the SDSS photometric galaxy sample has
a typical redshift of $z\sim 0.3$. However, 
there also appears too large
population of galaxies at $z\simlt 0.05$: since the redshift range
covers a very small volume, the galaxy population is very likely to be
largely contaminated by the outliers of photometric redshift estimates.
To avoid this contaminating population, we employ the lower cutoff
$z_{\rm ph, min}=0.1$ for our galaxy sample. 
Fig.~\ref{fig:photoz} also shows that there are not many galaxies beyond
$z = 0.9$. Furthermore, in order to study the impacts of low-redshift
nonlinear structures and galaxy magnitude cut, we study the four galaxy
catalogs listed in Table~\ref{tab:samp}. Note that the magnitude cut
difference in the bright and faint samples are intended to study a
possible contamination of imperfect galaxy-star separation.

\begin{table*}
\begin{center}
\begin{tabular}{l|cccccc}
\hline\hline
Galaxy Sample & 
\hspace*{1em}Magnitude range\hspace*{1em}&
$z_{\rm ph}$ range &
$\bar{n}_g$ [deg$^{-2}$] & 
$\bar{n}_g$ &
$\bar{n}_g$ &
$\bar{n}_g({\rm SGH})/\bar{n}_g({\rm NGH})$\\
& & &
(NGH) &
(SGH) &
(NGH+SGH) &\\ 
\hline
Bright-Shallow (BS) sample  & 
$19.1\le m_i \le 19.6$ & $0.1\le
      z_{\rm ph}\le 0.4$ 
&338.6
&329.1
&337.8
&0.972\\ 
Bright-Deep (BD) sample & 
$19.1\le m_i \le 19.6$ & $0.1\le
      z_{\rm ph}\le 0.9$ 
&502.4
&501.7
&502.3
&0.999
\\
Faint-Shallow (FS) sample   &$19.6\le m_i \le 20.1$ & $0.1\le
      z_{\rm ph}\le 0.4$ 
&453.1
&431.1
&451.2
&0.951
\\
Faint-Deep (FD) sample  &$19.6\le m_i \le 20.1$ & $0.1\le
      z_{\rm ph}\le 0.9$ 
&804.9
&774.8
&802.2
&0.963\\
\hline\hline
\end{tabular}
\end{center}
\caption{
The definition of the four galaxy samples used in the analysis of this
paper. The columns show the following:  
the $i$-band
 magnitude range imposed to define the galaxy sample; 
the photometric redshift range 
imposed to define the galaxy sample;
the average number density of galaxies for the Northern
 Galactic hemisphere (NGH) region with area 6928 deg$^2$;
the average number density for the Southern
 Galactic hemisphere (SGH) region with area 679 deg$^2$;
the average number density for the whole NGH+SGH
 region with area 7607 deg$^2$;
the ratio of the number densities of NGH and SGH regions. 
\label{tab:samp}
}
\end{table*}

\subsubsection{Constructing the galaxy number count field}
\label{sec:pixelization}

Our method described in Sec.~\ref{sec:Methodology} is applicable to
the pixelized data of galaxy distribution. For convenience, as done
in \cite{Yahata2007},  we employ
the same pixelization as that used in the SFD dust extinction map, which
is given in the format of pairs of 
4096$\times$4096 pixel Lambert projections, 
for each of the Northern and Southern Galactic
hemispheres (hereafter NGH and SGH, respectively). 
Note that the pixel size is $(2.\!\!{}^{\prime}372)^2$ (see Appendix~C
in SFD). 
The SDSS galaxy distribution has a partial sky coverage, so we do not 
use the pixels that are not 
included
in the 
survey region.
Also for safety
to avoid the pixelization effect, we do not use the pixels that reside
at boundaries of the SDSS survey regions.  For the remaining pixels,
dust extinction 
effect on 
galaxy magnitudes is corrected 
for based on the SFD map,
and then the galaxy number counts is computed in each pixel for each
galaxy sample. However, since we are interested in the dipole
(large-angular) anisotropy of galaxy distribution, we do not need
small-scale clustering information. Therefore for computational
convenience we create a coarser pixelized map of the galaxy counts in
which each pixel is defined by combining $73\times 73$ square-shape
neighboring pixels. The area of each pixel (without masking) is 8.34
deg$^2$.  Furthermore, the SDSS regions contain masked regions as
described before. Properly taking into account the masking effect, each
pixel is assigned to the area of unmasked region and the combined galaxy
number counts. Here for safety we do not use the pixels whose effective
area is smaller than 10\% of the pixel area after masking.  As a result,
the galaxy number counts we will work on are given in 983 and 168
pixels for the NGH and SGH regions, which have areas of 6928 and 679
deg$^2$, respectively (1148 pixels with 7607 deg$^2$ in total).

Table~\ref{tab:samp} summarizes the average number density for the four
galaxy catalogs. 
As mentioned in Sec.~\ref{subsubsec:reducing_a_clustering_dipole} above, 
we constructed the bright and faint samples to see effects of 
incomplete star-galaxy separation, expecting the faint samples suffer 
from more contamination. On the other hand, the shallow and deep samples
are prepared to see effects of the large scale structure contamination, 
expecting the shallow samples have more contaminant. Other than 
these
systematics, 
the average density of galaxies
is found to be larger in the NGH region than in the SGH region. Although
we are not sure for the significance due to the limited sky coverage of
SDSS survey, the constraints on the aberration effect is found to be
sensitive to an inclusion of the SGH region into the analysis as will be
shown below, because the number density difference, 
if it is not real, 
mimics the dipole
modulation of galaxy distribution.

\begin{center}
\begin{figure*}[t]
\centering\includegraphics[width=15cm]{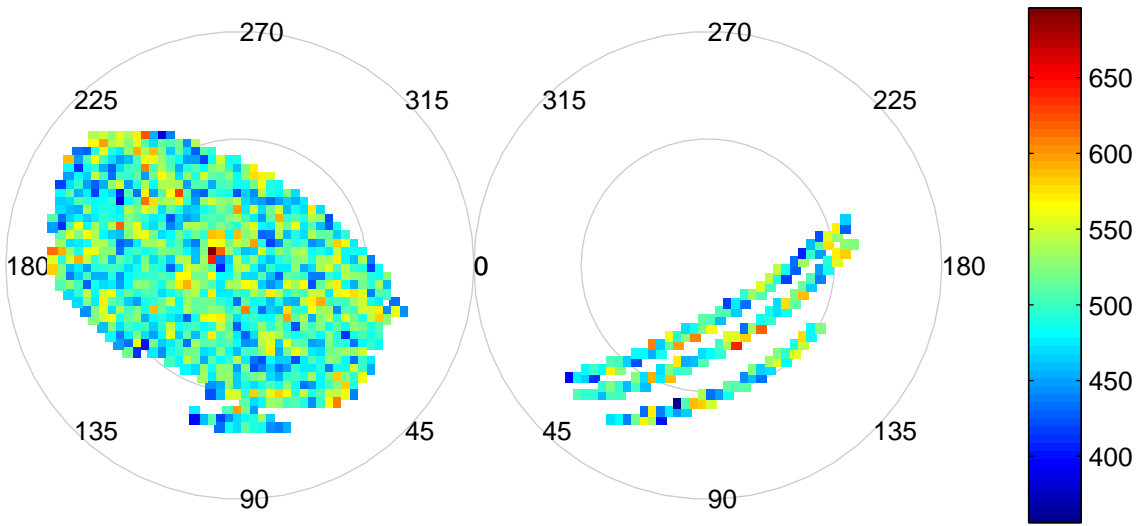}
\caption{The pixelized galaxy distribution of the BD sample in the Northern
 Galactic hemisphere (NGH: {\em left panel})
and the Southern Galactic hemisphere (SGH: {\em right}), respectively,
 based on the Lambert projection. 
The gray-color scales correspond to the number densities of galaxies in
each pixel per unit square degrees as indicated by the right-side bar.
In each panel the outer light-gray circle represents the zero galactic
latitude ($b=0^\circ$), while the inner circle $b=45^{\circ}$.  The
numbers labeled along the $b=0^{\circ}$ circle show the galactic
longitudes.
}  \label{fig:NumbercountSGH}
\end{figure*} 
\end{center}

Fig.~\ref{fig:NumbercountSGH} shows the pixelized galaxy distribution of
the BD sample in the NGH (left panel) and SGH (right), respectively. The
survey geometry can be clearly seen: the NGH region has a much greater
coverage than the SGH, and the SGH regions have three survey stripes.

\subsection{Computation of covariance matrix}
\label{sec:cov}

The $\chi^2$-estimation of the aberration effect, given by
Eq.~(\ref{eq:chi2definition}), requires an estimate of the covariance
matrix, ${\bf C}_{ij}$, where the indices
$i$ and $j$ run 
over the pixels of galaxy density map. 

As can be found from Eq.~(\ref{eqn:cov}) the diagonal components, ${\bf
C}_{ii}$, consist of two contributions: the Poisson noise arising due to
discreteness of galaxy distribution and the sampling variance arising
from the density fluctuations in 
large-scale structures. The Poisson
noise contribution can be directly computed from the galaxy number
counts in each pixel. On the other hand, the sample variance depends on
the variances of galaxy density fluctuations of pixel scales, and the
computation needs a few cares.  Firstly, the areas of 
pixels
are not
uniform due to masking, and may vary from $0.834$ to $8.34$ deg$^2$ as
described above.  Secondly, a theoretical estimate of the sample
variances (see Eqs.~\ref{eq:angularPowerSpectrum} and
\ref{eq:angularpowerspectrum}) involves several uncertainties: galaxy
bias uncertainty and nonlinear clustering uncertainties 
corresponding to the pixel scales. 
In other words the linear mass
fluctuations and the linear galaxy bias very likely break down on the
relevant scales.  Thus we instead estimate the sample variances directly
from the SDSS galaxy catalog itself.  
To do this we 
used the original
finest pixelized data of $(2.372)^2$ arcmin$^2$ in order to estimate
the variances as a function of the smoothing scales, where the smoothing
is done by combining neighboring pixels of square-shape region. Note that
the smoothing angular scale is simply estimated from the area of
combined pixels as 
$\theta_{\rm sm}=\sqrt{\Omega_{\rm pix}/\pi}$.

\begin{figure}
\begin{center}
\centering\includegraphics[width=9cm]{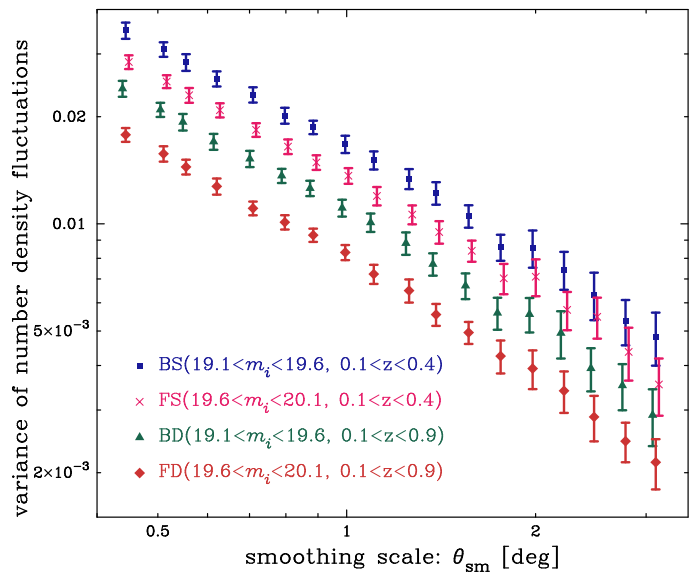} \caption{The
variances of galaxy number density fluctuations as a function of the
smoothing scales, for the four galaxy catalogs defined in
Table~\ref{tab:samp}. The variances are estimated from the pixelized
galaxy catalogs with varying pixel sizes (here 18 different-size pixels
are computed), and the smoothing scale in the horizontal axis is simply
estimated from the pixel area as $\theta_{\rm sm}=\sqrt{\Omega_{\rm
pix}/\pi}$ (see text for the details). Note that, for illustration
purpose, the results for the FS and BD samples are slightly shifted in
the horizontal direction.  The error bars around each data points are
estimated based on the bootstrap resampling method.  }
\label{fig:std_estimate_error}
\end{center} 
\end{figure} 

Fig.~\ref{fig:std_estimate_error} shows the estimated variances for the
four galaxy samples over the entire SDSS survey region, where the
variances are estimated for 18 different smoothing scales. We then
assign the sample variance of an arbitrary pixel area by 
spline-interpolating the 18 data points. Thus we
ignore the hole effect or pixel geometry 
due to masking for
simplicity. Since the galaxy sampling region within each pixel in the
galaxy catalog is not necessarily connected due to masking, the sample
variance estimated in our method arises from the density fluctuations of
the smallest smoothing scales for a fixed area and would be greater than
the actual sample variance. Thus our estimate of the sample variance is
somewhat conservative.

Recall that, as given by Eq.~(\ref{eqn:cov}), the Poisson noise
contribution to the covariance is given by $1/(\bar{n}_g\Omega_{{\rm
pix}(i)})$ for a pixel with area $\Omega_{{\rm pix}(i)}$ ($\bar{n}$ is
the average number density), where $\Omega_{{\rm pix}(i)}$ varies from
0.834 to 8.34 deg$^2$.  Table~\ref{tab:samp} tells that, for our galaxy
catalogs, the Poisson noise is in the range of
$O(10^{-4})-O(10^{-3})$. On the other hand, since the pixel areas in the
range of 0.834 to 8.34 deg$^2$ correspond to the smoothing scales from
0.52 to 1.63 deg in Fig.~\ref{fig:std_estimate_error}, the sample
variance is found to be 
$O(10^{-2})$.
Therefore, for our galaxy catalogs, the sample variance dominates over
the Poisson noise in the diagonal covariance components, more then by a
factor 10, i.e. (samp. vari.)$>$10 (Poisson noise), thanks to the enormous
size of  SDSS galaxy catalog.

An estimate of the off-diagonal covariance components, ${\bf C}_{ij}
(i\ne j)$, is more straightforward, where it contains only the sample
covariance contribution. Since different pixels of our galaxy catalogs
are separated by more than 2.89($\approx \sqrt{8.34}$) degrees, the
cross-correlation between the galaxy density fluctuations of different
pixels are safely considered to be in the linear regime. Therefore we
use 
Eq.~(\ref{eq:angularPowerSpectrum})
to estimate the off-diagonal covariances. The
computation requires several ingredients. For redshift projection, we
used the photometric redshift distribution of galaxies in
Fig.~\ref{fig:photoz}. As for the linear
mass power spectrum we employed the transfer function given in 
\cite{Eisenstein1999} 
assuming the WMAP-3year cosmology \citep{Spergel2007}. As can be found
from 
Eq.~(\ref{eq:angularPowerSpectrum})
the pixel window function is needed to specify; 
we adopted a top-hat type window function $W(\bi{\theta})=1/(\pi
\theta_{\rm th}^2)$ where $\theta_{\rm th}$ is simply estimated from the
pixel area as $\theta_{\rm th}=\sqrt{\Omega_{{\rm
pix}(i)}/\pi}$ \footnote{Here we simply ignore the square pixel shape and
also the masking effect within one pixel. However, our approximation is
good enough for the large-angle cross-correlations.}.
Finally we also need to specify the linear bias parameter of each galaxy
sample, $b_g$. We estimated $b_g$ by fitting the variances 
at large angular scales $1.5^{\circ} \le \theta \le
2.6^{\circ}$ to the linear theory predictions, 
resulting in the linear bias
parameters $b_g \simeq 1.0$ for both the BS and FS sample galaxies,
while $b_g \simeq 1.2$ for the BD and FD samples.

\subsection{Expectation}
\label{sec:SDSS_DR6_Expectation}

\begin{center}
\begin{figure}
\centering\includegraphics[width=13cm]{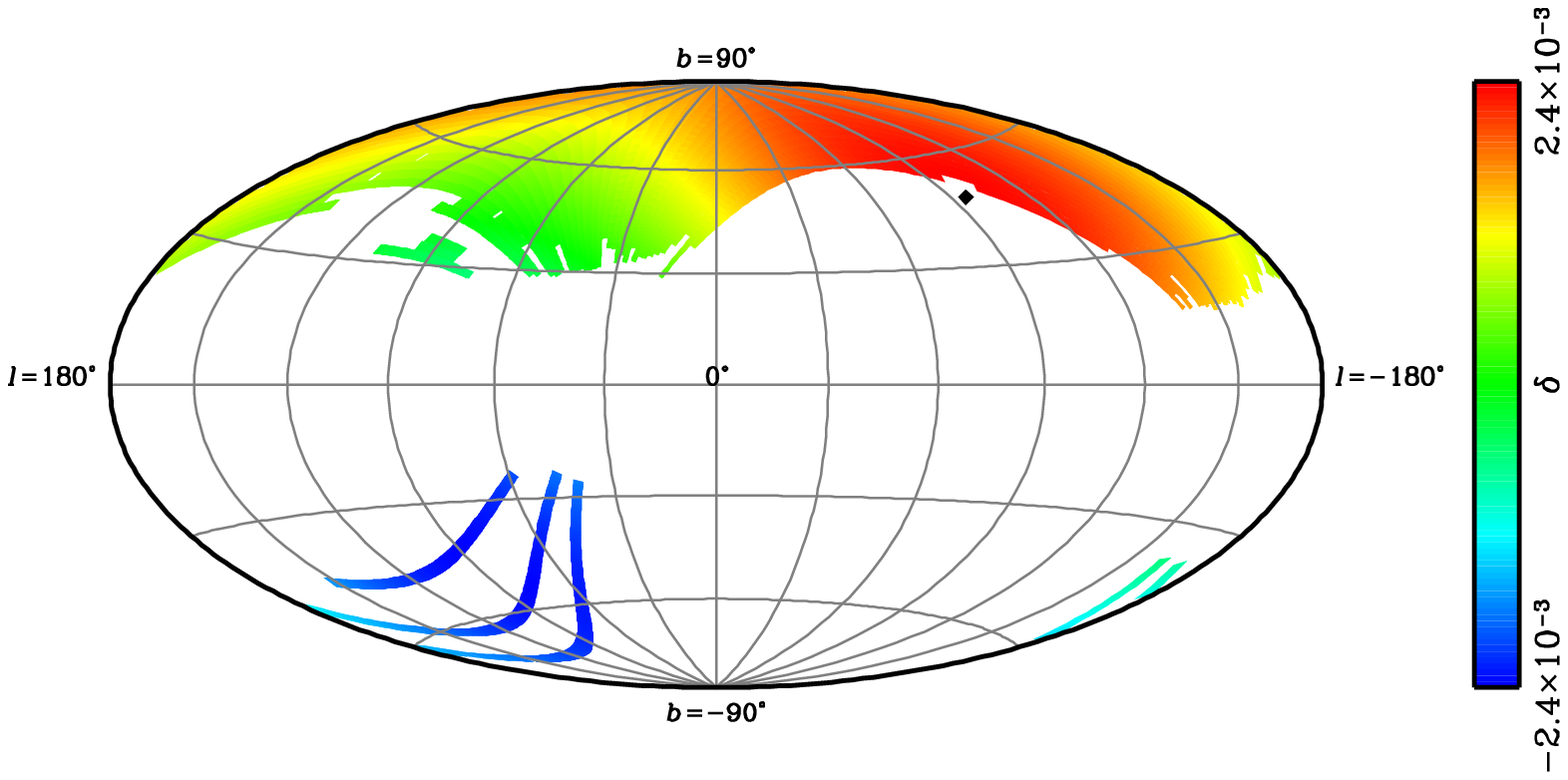}
\centering\includegraphics[width=11cm]{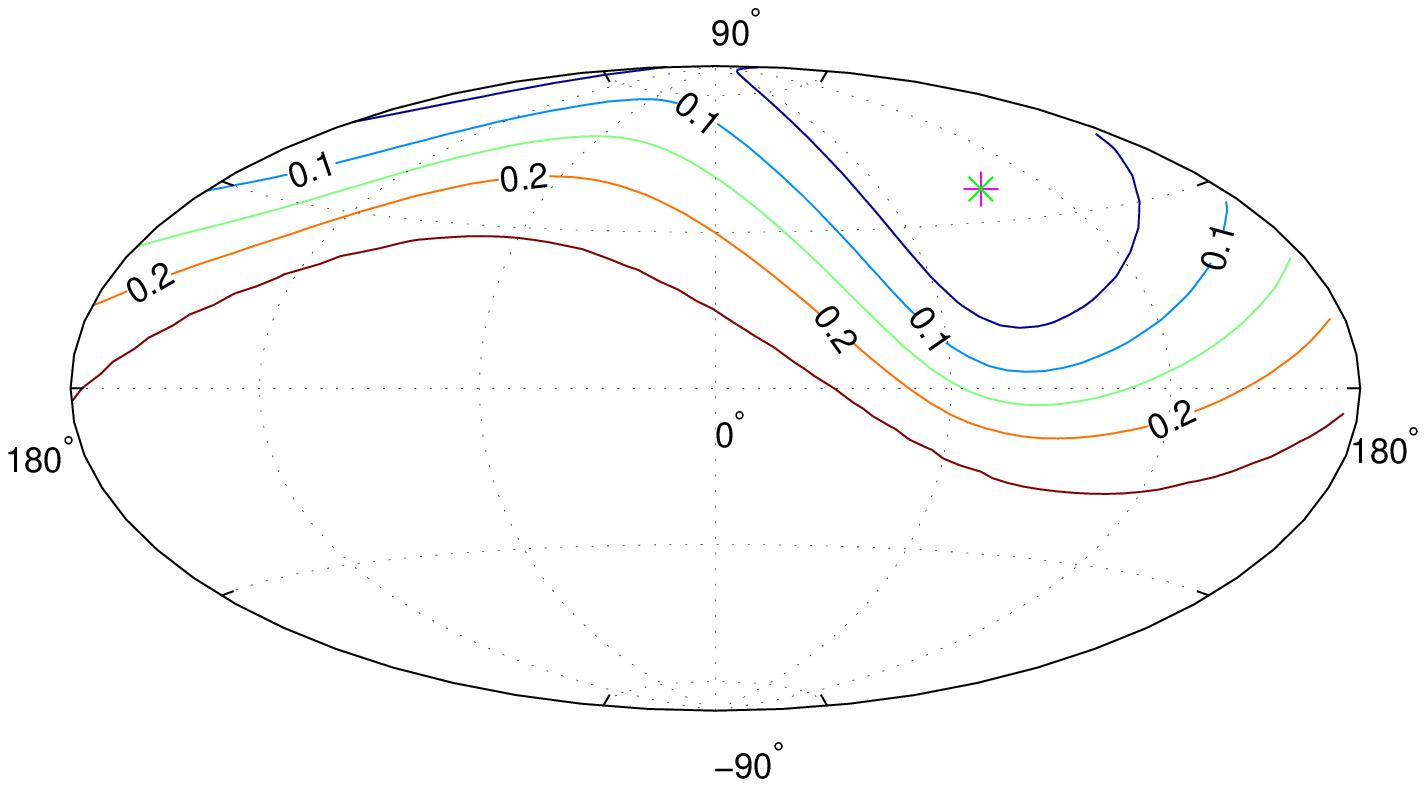}
\caption{{\em Upper panel}: An expected dipole pattern of the galaxies
number density field for the SDSS survey region. Here
we assumed the dipole
amplitude and direction as those of the CMB dipole: the
velocity amplitude $\beta=1.231\times 10^{-3}$, and the angular direction
is denoted by the black square symbol. 
We set $\tilde \beta = \beta$ in 
Eq.~(\ref{eqn:tilde_beta}) for simplicity.
The color scales denote the
number density fluctuations of galaxies as indicated by the right-hand
color bar. {\em Lower panel}: The expected accuracy of constraining the
dipole angular direction by the $\chi^2$ fitting
(Eq.~\ref{eq:chi2definition} or
 ~\ref{eq:theoreticallyExpectedDeltaChi2}) , 
where the covariance matrix for the FD
sample is used as a representative example. The contours denote the
$\langle \Delta\chi^2 \rangle $ 
distribution over the sky, and show that the angular
direction is not constrained: all the directions are within the
$1\sigma$ region ($\langle\Delta\chi^2\rangle\le 2.3$ 
for two parameter case). 
The cross and the plus symbols indicate the $\chi^2$ minimum and 
the CMB dipole direction, respectively.  
The $\langle\Delta \chi^2\rangle$ value corresponding to each contour line 
is indicated by the numbers on the line.
}
\label{fig:SDSS_expected_dipole_pattern_s.eps}
\end{figure} 
\end{center} 

Before going to the results, the upper panel of
Fig. \ref{fig:SDSS_expected_dipole_pattern_s.eps} shows the expected
dipole modulation pattern of galaxy distribution in the SDSS survey
region, {\em if} the aberration effect on the SDSS galaxy distribution
is in the same amplitudes and angular direction as in the CMB dipole,
i.e. if the Earth's peculiar velocity to the SDSS galaxy distribution is
$\beta_{\rm CMB}=1.231\times 10^{-3}$ in the direction 
$(l_{\rm
CMB},b_{\rm CMB})=(264^\circ\!\!.31,48^\circ\!\!.05)$ in the Galactic
coordinates.
For simplicity we ignore the Doppler effect,
i.e., set $\tilde \beta = \beta$ in 
Eq.~(\ref{eqn:tilde_beta}).
The plot shows the $O(10^{-3})$-level density modulation
can be expected, however, the partial sky coverage of the SDSS region seems  
to significantly obscure the characteristic pattern and prevent the
detection. It may also be worth noting that the CMB dipole direction
predicts a negative number density contrast in the SGH region, which may
be indicated from our galaxy catalogs as shown in Table~\ref{tab:samp},
although it may be just a  coincidence.

\begin{center}
\begin{figure}
\centering\includegraphics[width=10cm,height=6cm]{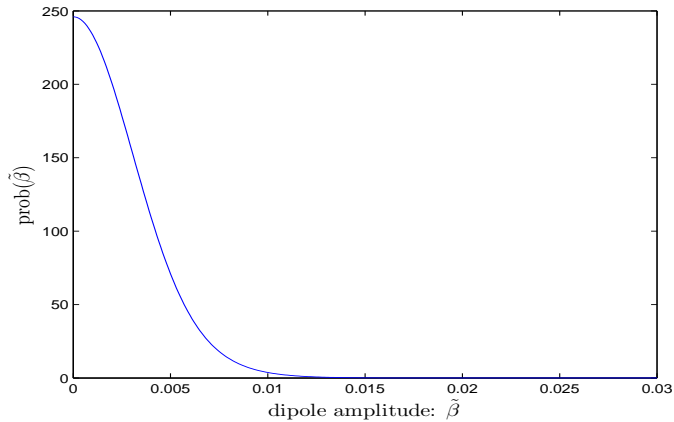}
\caption{An expected marginalized likelihood of the dipole amplitude
$\tilde \beta$ is shown for the SDSS galaxy distribution depicted in
Fig.~\ref{fig:SDSS_expected_dipole_pattern_s.eps}. The SDSS catalog
allows for an accuracy of $\tilde \beta=O(10^{-2})$ for the aberration
 effect search.} \label{fig:prob_vel_10z90_196i201_exp}
\end{figure} 
\end{center} 

Assuming the 
dipole density modulation shown in the upper panel of 
Fig. \ref{fig:SDSS_expected_dipole_pattern_s.eps}, we then compute the
mean $\chi^2$ values (minus the degrees of freedom, or ``d.o.f.'') with varying free 
parameters $(\hat{\bfm{v}}, \tilde \beta)$. 
In other words, using Eq. (\ref{eq:chi2definition}) 
we compute 
 \begin{eqnarray}
\lefteqn{\langle\Delta \chi^2(\hat{\bfm{v}}, \tilde \beta)\rangle \equiv 
\langle\chi^2(\hat{\bfm{v}}, \tilde \beta)\rangle 
- ({\rm d.o.f}) 
} \nonumber \\
&=& \sum_{i,j=1}^{N_{\rm pix}}
\left[
\langle\delta_{\rm assumed}(\bi{\theta}_i)\rangle-\delta_{\rm model}(\bi{\theta}_i; \hat{\bfm{v}},\tilde \beta)
\right]
\left[\bfm{C}\right]^{-1}_{ij}
\left[
\langle\delta_{\rm assumed}(\bi{\theta}_j)\rangle-\delta_{\rm
model}(\bi{\theta}_j;\hat{\bfm{v}}, \tilde \beta)
\right],
\label{eq:theoreticallyExpectedDeltaChi2}
\end{eqnarray}
where $\langle f\rangle$ denotes an ensemble average of 
the 
quantity $f$ and 
\begin{eqnarray}
\langle\delta_{\rm assumed}(\bi{\theta}_j)\rangle &=& 2 
\beta_{\rm CMB} \hat{\bfm{v}}_{\rm CMB} \cdot{\bi{\theta}}_j. 
\end{eqnarray}
(By 
subtracting
the degrees of freedom, this quantity becomes zero when our free parameters 
$(\hat{\bfm{v}}, \tilde \beta)$ coincide with the
hypothetical observation parameters
$(\hat{\bfm{v}}_{\rm CMB},
\beta_{\rm CMB})$ in this case).
The survey geometry of the SDSS DR6 are taken into
account when computing $\bi{\theta_j}$, the covariance matrix and $\delta_{\rm model}$.

The lower panel of
Fig. \ref{fig:SDSS_expected_dipole_pattern_s.eps} then 
gives a more quantitative estimate, showing the accuracy
of determining the dipole angular direction from the expected number
density modulation. To be more explicit, this plot shows the
theoretically expected mean $\chi^2$
difference between the best-fit model and models with 
varying model parameters to specify the angular
direction of the peculiar velocity, marginalized over the velocity
amplitude 
computed from Eq. (\ref{eq:theoreticallyExpectedDeltaChi2}): 
$\langle\Delta\chi^2(l,b)\rangle$. 
We used the covariance matrix
for the FD galaxy sample (see Sec.~\ref{sec:cov}) in order to compute
the $\Delta\chi^2$.
The figure clearly shows that $\langle\Delta\chi^2\rangle$
is smaller than 1 over the
entire sky. That is, the aberration effect is very difficult to detect,
if the peculiar velocity to the SDSS galaxy is similar to the CMB dipole
amplitude. In other words, the covariance, especially the intrinsic
galaxy clustering contamination, is so significant compared to the
aberration effect. This result is consistent with a rough estimate shown
in Table~\ref{tab:sn}. Fig.~\ref{fig:prob_vel_10z90_196i201_exp} shows
the marginalized probability distribution of the dipole 
 amplitude parameter
$\tilde \beta$. Only an upper limit on $\tilde \beta$ at the level
$\tilde \beta\simlt
10^{-2}$ is likely to be obtained from the SDSS catalog.

\subsection{Results}
\label{sec:SDSS_DR6_Results}

\begin{table}
\begin{center}
\begin{tabular}{l|clccc}
\hline\hline 
Galaxy Sample     &  \hspace*{0.5em}$\chi^2_{\rm min}$ \hspace*{0.5em}
& best-fit model: ($\tilde \beta,l,b$)\hspace*{0.5em}
& Errors: ($\Delta \tilde \beta$, $\Delta \theta$) 
&  \hspace*{0.5em}$\chi^2_{\rm CMB} -\chi^2_{\rm min}$ \hspace*{0.5em}
&  \hspace*{0.5em}$\chi^2_{\beta=0}-\chi^2_{\rm min}$ 
\\ \hline 
BS sample    & 1080.3 
& (0.0127,310$^{\circ}$,30$^{\circ}$)
&  $\left(\raisebox{0.8ex}{+0.0067}\hspace{-3.4em}\raisebox{-1.0ex}{--\hspace{0.1em}0.0054}, 64^{\circ}\right)$ 
& 5.85  & 7.22  \\ 
BD sample    & 1211.4 
& (0.0087,290$^{\circ}$,$-10^{\circ}$)
&  $\left(\raisebox{0.8ex}{+0.0059}\hspace{-3.4em}\raisebox{-1.0ex}{--\hspace{0.1em}0.0057}, 100^{\circ}\right)$ 
& 2.01  & 2.35  \\
FS sample    & 1108.0 
& (0.0156,290$^{\circ}$,60$^{\circ}$)
&  $\left(\raisebox{0.8ex}{+0.0039}\hspace{-3.4em}\raisebox{-1.0ex}{--\hspace{0.1em}0.0034}, 40^{\circ}\right)$ 
& 18.7 & 21.7  \\ 
FD sample    & 1288.1 
& (0.0121,280$^{\circ}$,75$^{\circ}$) 
& ($\pm0.0023$, 33$^{\circ}$) 
& 21.2 & 25.6   \\
\hline \hline 
\end{tabular} 
\caption{The results for the $\chi^2$ fitting for the four SDSS galaxy
samples defined in Table~\ref{tab:samp}. The second column labeled as
``$\chi^2_{\rm min}$'' denotes the minimum $\chi^2$ value for the
best-fit model, and the third column gives the best-fit parameters. 
The fourth column shows, for each sample, 
the $1\sigma$ error
for $\tilde
 \beta$ and 
the $1\sigma$ uncertainty of determining the dipole direction in 
radius from the best-fitting angular position. 
The
fifth column gives the $\chi^2$-difference between the best-fit model and the
case that 
the peculiar motion induced dipole has  
the same amplitude and angular direction as those of the CMB dipole, 
given as $(\beta_{\rm CMB},l_{\rm CMB},b_{\rm CMB})
= (1.231\times10^{-3},264.\!\!^{\circ}31,48.\!\!^{\circ}05)$.
The sixth 
column shows the $\chi^2$-difference compared to no aberration effect
case, $\beta=0$.} 
\label{tbl:SDSS_chi2result}
\end{center} 
\end{table} 

Now let us move on to the measurement results. 
Here in the analysis on the real data, we first count number of galaxies in
each pixel and compute the density perturbation field using 
Eq. (\ref{eqn:delta_obs}). Second, choosing model parameters
$(\hat{\bfm{v}},\tilde \beta)$, we compute $\delta_{\rm model}$ using 
Eq. (\ref{eqn:delta_model}) and construct 
$\chi^2(\hat{\bfm{v}},\tilde \beta)$ using Eq. ~(\ref{eq:chi2definition}) 
for those particular model
parameters. Varying the model parameters, we iterate the second step, 
searching 
for the parameters that minimize the
$\chi^2$ value. 
The value of $\tilde{\beta}$ is 
varied in the range of $0 \le \tilde \beta \le 7.0\times 10^{-2}$ and
the values of ($l,b$) are varied over the entire sky directions.
We used the covariance matrix computed based on the method in
Sec.~\ref{sec:cov} for each galaxy sample. 

Table~\ref{tbl:SDSS_chi2result} summarizes the $\chi^2$-fitting results
for the four galaxy samples listed in Table~\ref{tab:samp}. 
The second- and third-columns in the table give the minimum $\chi^2$
value, $\chi^2_{\rm min}$, for the best-fit model and 
its model parameters.
The fact $\chi^2\approx (\mbox{pixel number})$ implies that 
our covariance matrix estimate is reasonable.
The table shows that the best-fit values of $\tilde{\beta}$ are of the
order of $O(10^{-2})$ as expected from
Fig.~\ref{fig:prob_vel_10z90_196i201_exp}. 
Table~\ref{tbl:SDSS_chi2result} also shows the $\chi^2$-differences,
$\Delta\chi^2$, obtained by comparing the $\chi_{\rm min}^2$ with two
models: one assumes the CMB dipole amplitude and direction for the
aberration effect, and the other is no aberration effect
($\beta=0$). For both cases, the $\chi^2$ differences are modestly
large, except for the BD sample, implying that the dipole angular
modulation pattern, which is not necessarily the aberration effect, is
marginally detected.

\begin{center}
\begin{figure}
\centering\includegraphics[width=12cm]{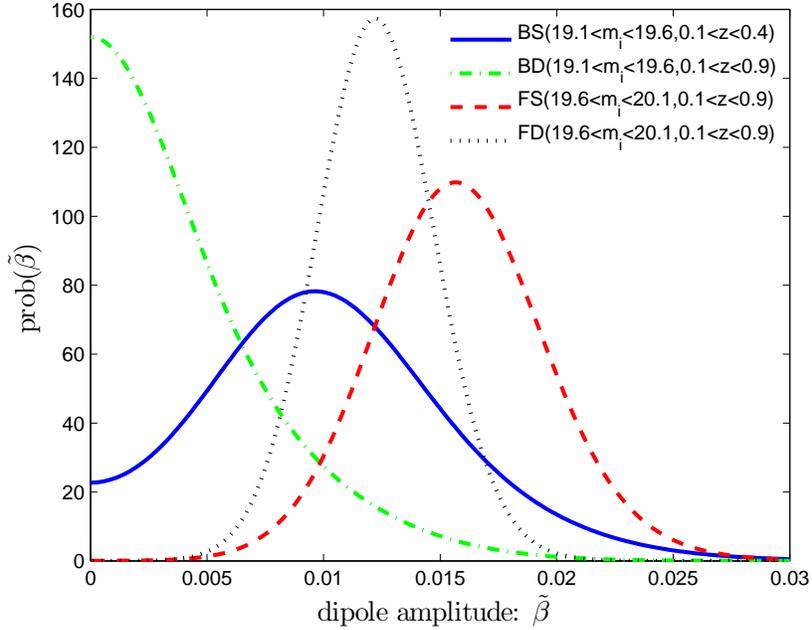}
\caption{The likelihood probability distribution for the 
dipole amplitude parameter $\tilde \beta$ of the aberration effect,
marginalized over the angular direction parameters $(l,b)$, for the four
galaxy samples. The FS and FD samples which preferentially include fainter
galaxies 
prefer a non-zero $\tilde \beta$ with the amplitudes
$O(10^{-2})$. The other two samples are consistent with zero $\tilde \beta$. }
\label{fig:prob_beta}
\end{figure} 
\end{center} 

\begin{center}
\begin{figure}
\centering\includegraphics[width=12cm]{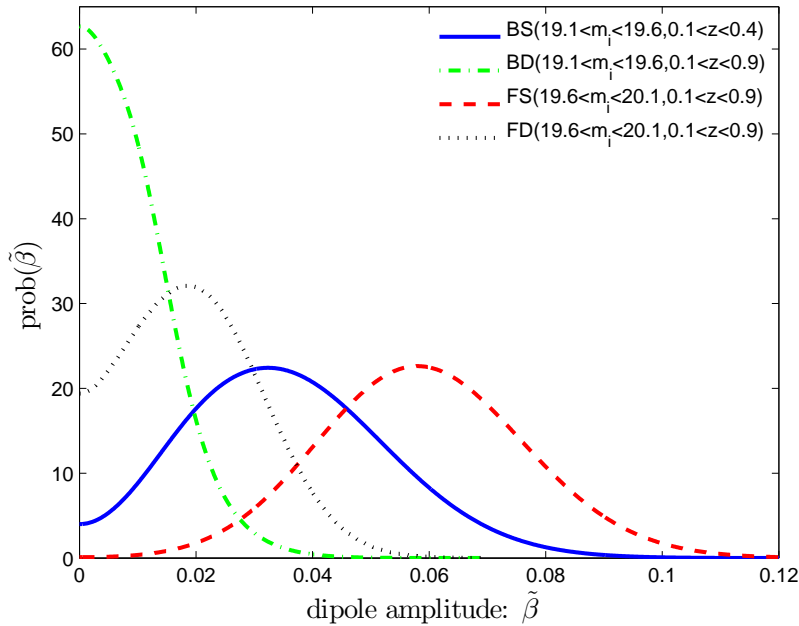}
\caption{ Estimated marginalized probability of the dipole amplitude
$\tilde \beta$, where only
 the data in the Northern Galactic hemisphere region is used.   
} \label{fig:prob_beta_ngh}
\end{figure} 
\end{center}

Fig.~\ref{fig:prob_beta} is the marginalized probability distribution of
the dipole amplitude parameter $\tilde \beta$
for the four galaxy
samples. The FS and FD samples seem to prefer non-zero $\tilde \beta$, however,
the likely amplitude is $O(10^{-2})$, implying that the peculiar
velocity of the Earth relative to the SDSS galaxy distribution is  $v\sim
3000\mbox{km s}^{-1}$. Such a large peculiar velocity is difficult to
explain in the currently concordance $\Lambda$CDM model, and therefore
we believe this is likely due to either the intrinsic galaxy clustering
contamination or 
unresolved
systematic effects, or both, as also demonstrated in
Figs.~\ref{fig:SDSS_expected_dipole_pattern_s.eps} and
\ref{fig:prob_vel_10z90_196i201_exp}. We will come back to this 
issue later in this section.

\begin{center}
\begin{figure}
\centering\includegraphics[width=14cm]{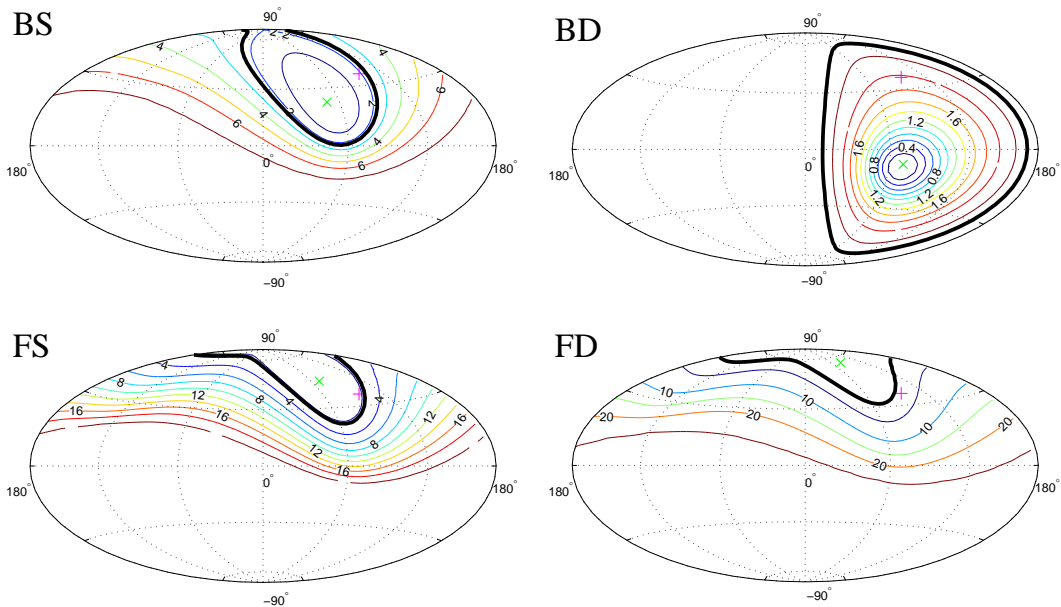}
\caption{The $\Delta \chi^2$ distribution as a function of the dipole
direction parameters $(l,b)$ on the sky, marginalized over the dipole 
 amplitude parameter $\tilde \beta$. The upper-left, upper-right,
lower-left and lower-right panels are the results for the BS, BD, FS,
and FD samples, respectively.  The thick (black) contour corresponds to
the $1\sigma$ uncertainty region 
($\Delta \chi^2 = 2.3$). The
`$\times$' symbol denotes the best-fit direction,
 while the `$+$' symbol
denotes the CMB dipole direction, 
$(l_{\rm CMB},b_{\rm
CMB})=(264.\!\!^\circ31, 48^\circ\!\!.05)$.  
}
\label{fig:deltachi2_map}
\end{figure} 
\end{center} 

Fig.~\ref{fig:deltachi2_map} shows how the $\Delta\chi^2$ value,
marginalized over the dipole amplitude parameter $\tilde \beta$, varies as a
function of the peculiar velocity directions on the sky. It would be
interesting to find that the method developed here can constrain the
velocity directions. Again, for the FS and FD cases, the dipole directions
are constrained at more than $3\sigma$ level ($\Delta\chi^2$ = 11.8). 
Also interestingly, the preferred dipole direction is close to the CMB
dipole direction. If a possible detection of the galaxy dipole is due to
the intrinsic clustering contamination, the inferred direction is not
necessarily similar to the CMB direction 
since we have removed the contamination due to the nearby large scale structures 
using the photometric redshift information.

Now let us discuss possible causes of the indicated dipoles in our 
samples other than the BD sample. 
We construct the four galaxy catalogs to study 
possible remaining systematic effects due to the star-galaxy separation
and the intrinsic clustering contamination, as are defined in Table
\ref{tab:samp}.  As shown in Table~\ref{tbl:SDSS_chi2result} and
Fig.~\ref{fig:deltachi2_map}, the ``faint'' samples, which are supposed to
be more contaminated by an imperfect star-galaxy separation than for the
``bright'' samples, tend to prefer the dipole directions at higher
galactic latitudes. This is counter-intuitive, because the faint samples
may suffer from more star contamination 
towards lower galactic latitudes, which in turn causes 
the dipole direction in the lower
latitudes due to the enhanced overdensity. Hence the imperfect star-galaxy
separation seems not a main source of the systematic effects. 
On the other hand, if comparing
the ``deep'' and ``shallow'' samples,
the shallower sample is
supposed to be more affected by the intrinsic galaxy clustering
contamination from more evolving nonlinear structures at lower
redshifts, 
if the photometric redshifts are well reliable. 
In fact Fig. \ref{fig:prob_beta} shows that the dipole
amplitudes tend slightly larger for shallower samples. 

\begin{center}
\begin{figure}
\centering\includegraphics[width=14cm]{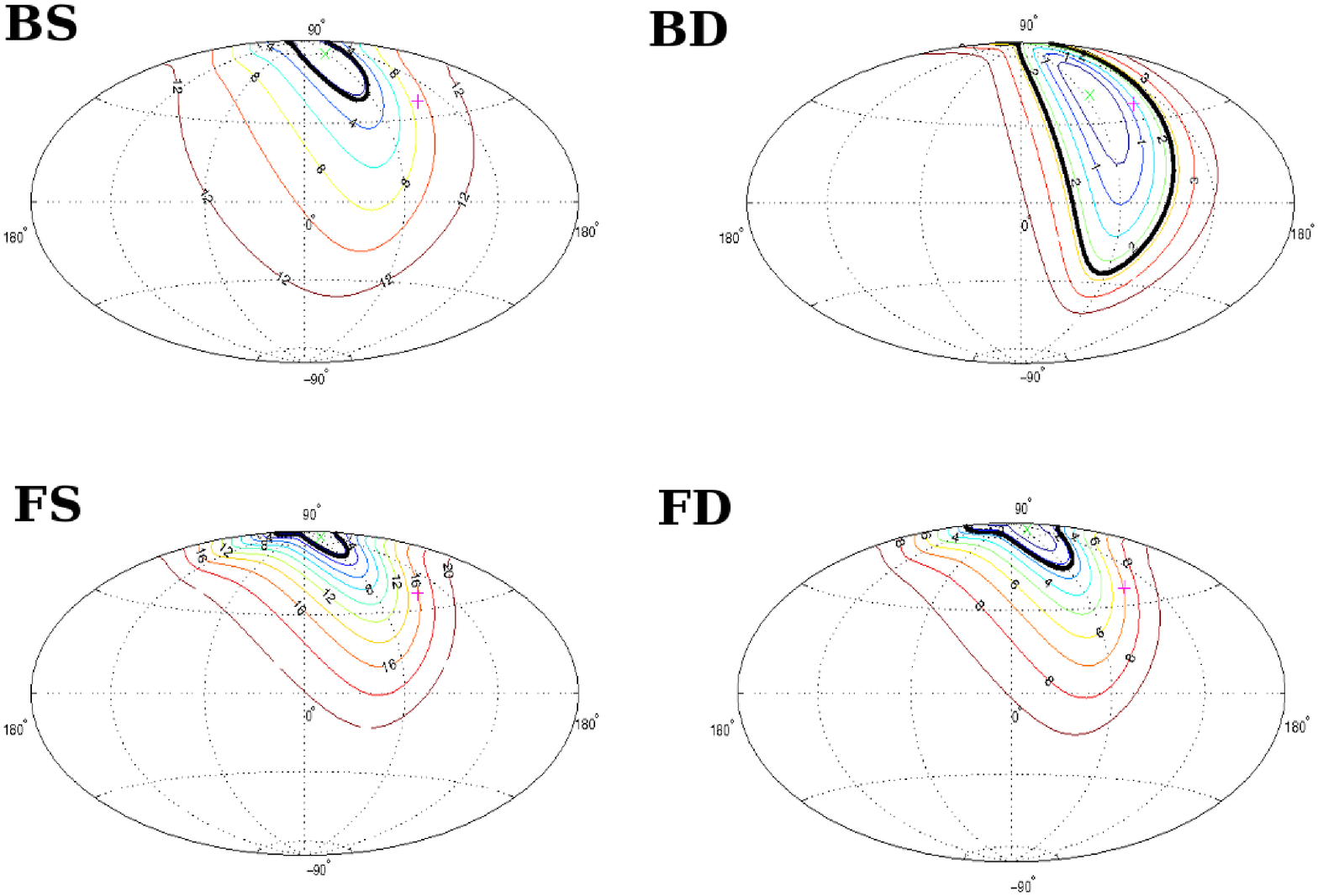}
\caption{
The same as Fig. \ref{fig:deltachi2_map}, but 
here for these plots, we have used the SDSS DR6 data in the Northern Galactic hemisphere only.
}
\label{fig:deltachi2_map_NGHonly}
\end{figure} 
\end{center}

Another concern arises from the
partial sky coverage. In particular, as discussed in
Table~\ref{tab:samp}, the SDSS galaxy samples tend to have a smaller
average number density in the Southern Galactic hemisphere than in the
Northern Galactic hemisphere, which may cause an (apparent) dipole
anisotropy. Fig.~\ref{fig:prob_beta_ngh} shows the probability
distribution of $\tilde \beta$ if only the NGH region is used in the
analysis. Now all four samples show broader distributions, and all but FS 
sample are consistent with zero $\tilde \beta$.
Fig. \ref{fig:deltachi2_map_NGHonly} 
shows the same plots as
Fig. \ref{fig:deltachi2_map}, 
but here again we have excluded the data in the Southern Galactic hemisphere 
from our four samples in the analysis. 
The $1\sigma$ contours in the four plots become
smaller, indicating larger dipole amplitudes as consistent with 
the finding on Fig.~\ref{fig:prob_beta_ngh} above.  
We also note that the 
minimum $\chi^2$ 
values
of all four samples (green crosses in
Fig. \ref{fig:deltachi2_map_NGHonly})  
more or less tend to move toward the Northern Galactic pole than in 
Figs.~\ref{fig:deltachi2_map}.  
Hence the number density differences between NGH and SGH
does move the direction of the detected dipoles in all the four samples toward
the CMB dipole direction, yet the origin of the difference itself is
unknown. As far as direction is concerned, the aberration induced
dipole expected from the CMB dipole does predict such a difference
between NGH and SGH as is indicated in the upper panel of 
Fig. \ref{fig:SDSS_expected_dipole_pattern_s.eps}. 
A data having a wider sky coverage, such as LSST,
is desired to settle down this issue.

The reason for all but BD samples, when excluding the SGH data,
show 
the minimum $\chi^2$
values
around the
Northern Galactic pole may be attributed to imperfect Galactic
dust extinction correction, although it is not perfectly clear 
how Galactic dust can produce the
difference in the BS and BD samples (or, in other words, how effects of 
Galactic dust differ among different redshifts).  
In this paper we have assumed that the dust extinction correction by the SFD dust map is perfect, but it is
well-known in the literature (see, e.g., \cite{Yahata2007}) that there
may remain
 some residual systematics in the map. 
In summary,
since 
the BD sample with and without the data in the SGH, 
which is the most robust sample against possible systematics 
among our four samples, 
indicates zero dipole, we conclude that the aberration effect is not
detected 
from the SDSS DR6 in our analysis.

\section{Prospect for a future wide-area survey: LSST}
\label{sec:FutureProspect}

As we have so far shown, a survey with large sky coverage and
sufficiently deep redshift coverage is required to detect the aberration
effect. The planned galaxy survey, the Large Synoptic Survey Telescope
(LSST)~\citep[][see also http://www.lsst.org/]{LSST}, will provide us with
a most promising opportunity to explore the aberration effect. Here we
estimate the expected accuracy.

\begin{center}
\begin{figure}
\centering\includegraphics[width=16cm]{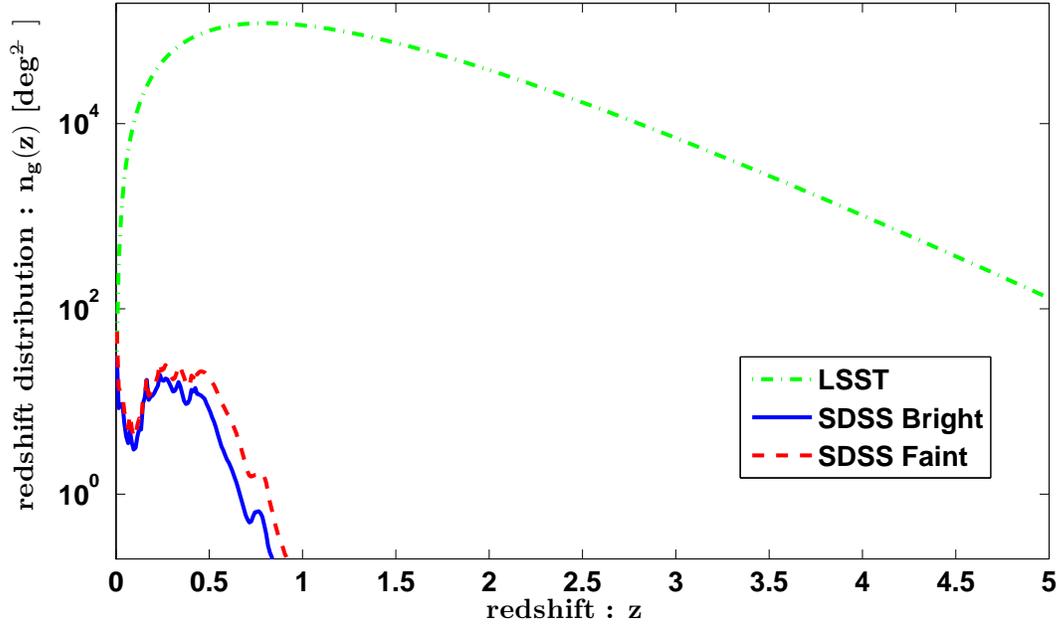}
\caption{
The expected redshift distribution $n_g(z)$ of the LSST survey
 (dash-dotted line).
The mean angular number density of 50 per unit square arcminutes is
 assumed \cite{Huterer2006}. Also shown for comparison are the galaxy redshift distribution of
 the SDSS bright sample $19.1 < m_i < 19.6$ (solid line) and of 
 the SDSS faint sample  $19.6 < m_i < 20.1$ (dashed line).
}
\label{fig:redsfhitdist_LSST}
\end{figure} 
\end{center}

To estimate the forecasts, we need to specify survey parameters.
According to \citep{Ivezic2006}, the survey area is assumed to be 18863
deg$^2$, and the survey geometry is restricted to the region
$-75^{\circ} \le \delta \le 15^{\circ}$, excluding the galactic disk
region of $-15^{\circ} \le b \le 15^{\circ}$ (also see the figure
below). 
For the galaxy redshift distribution, we simply assume the
analytic form given by Eq.~(4) in  \cite{Huterer2006}: 
\begin{eqnarray}
n_g(z) &\propto& z^2{\rm e}^{-z/z_0}
\label{eq:nofz}
\end{eqnarray}
with $z_0 = 0.4$, which has the
mean redshift $\bar{z}=1.2$. The galaxy distribution is normalized so as
to have the mean number density of galaxies of $50$ per unit square
arcminutes \cite{Huterer2006}. The assumed LSST galaxy redshift distribution is
shown in Fig. \ref{fig:redsfhitdist_LSST} along with the ones for the
SDSS samples shown in Fig. \ref{fig:photoz}.

\begin{center}
\begin{figure}
\centering\includegraphics[width=16cm]{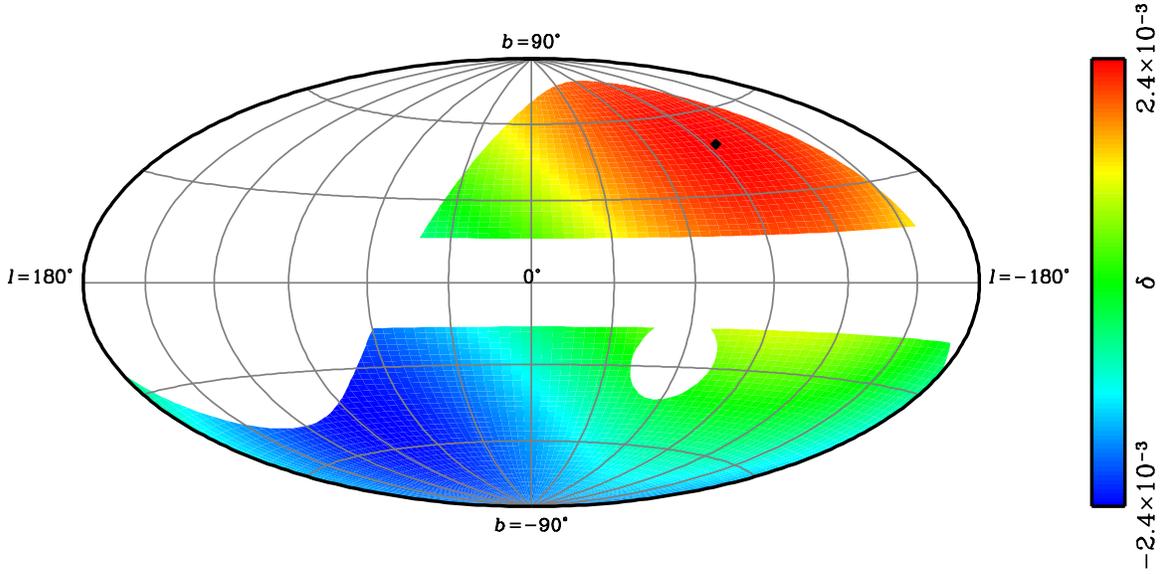}
\caption{The expected dipole pattern of the galaxies number density
 field 
expected
for 
a hypothetical 
LSST-like survey, 
assuming the CDM dipole amplitude and direction as in 
Fig.~\ref{fig:SDSS_expected_dipole_pattern_s.eps}.  
The diamond symbol indicates the
 CMB dipole direction.
}
\label{fig:LSST_exp}
\end{figure} 
\end{center} 

Fig.~\ref{fig:LSST_exp} shows the expected
dipole anisotropy pattern in the LSST galaxy distribution, assuming the
CMB dipole amplitude and directions. 
(For simplicity, we ignore the Doppler effect,
i.e., 
set $\tilde \beta = \beta$
in 
Eq.~(\ref{eqn:tilde_beta}) to produce the figures in
this section.)
It is clear that the wide
sky-coverage can nicely capture the dipole pattern.

\begin{center}
\begin{figure}
\centering\includegraphics[width=10cm]{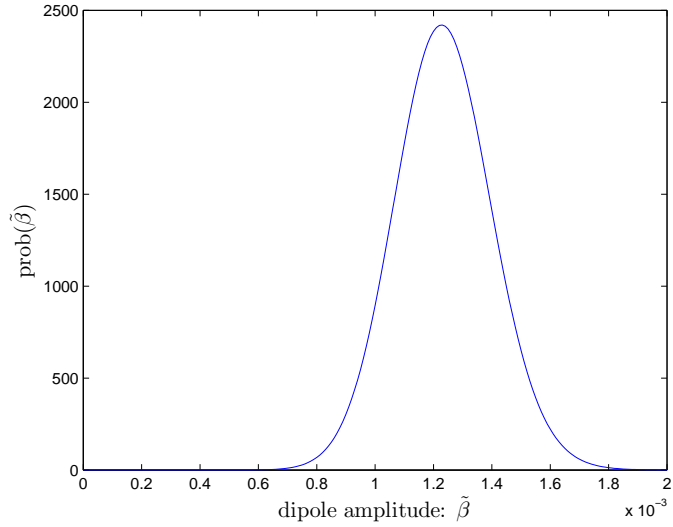}
\caption{The expected marginalized probability of the dipole amplitude
 parameter  $\tilde \beta$ for the aberration effect on the LSST galaxy
 distribution in Fig.~\ref{fig:LSST_exp}. 
}
\label{fig:prob_beta_lsst}
\end{figure} 
\end{center} 
\begin{center}
\begin{figure}
\centering\includegraphics[width=10cm]{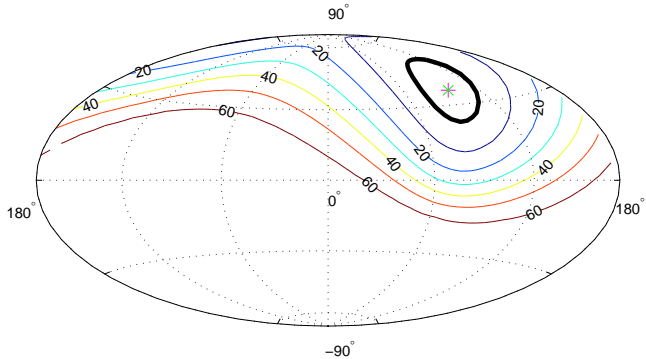}
\caption{The expected mean 
$\Delta \chi^2$ (i.e., $\langle\Delta\chi^2\rangle$ given by 
Eq. (\ref{eq:theoreticallyExpectedDeltaChi2})) 
distribution for the LSST survey.
The contours are stepped by $\langle\Delta\chi^2\rangle=10$, 
while the thick contour
denotes the $1\sigma$ region. The $1\sigma$ accuracy of determining the
angular dipole direction is $\sim 20$ degrees in radius.  }
\label{fig:deltachi2_lsst}
\end{figure} 
\end{center} 

To quantify the detectability of the aberration effect with the LSST, we need
to model the sample variance contribution to the covariance matrix. We
first considered the pixelized map with 213 pixels; each pixel has an
area of $88$ square degrees or $\sim 10$ degrees scale. The pixel size is
sufficiently large, so we use the linear theory to compute the sample
variance contributions. However, to reduce the intrinsic
clustering contamination from structures at low redshifts, we only
include galaxies at redshifts greater than $z=1$. Also we assume the
linear bias parameter $b_g=1$ for simplicity.

Figs.~\ref{fig:prob_beta_lsst} and \ref{fig:deltachi2_lsst} show the
forecasts, which show the marginalized probability of the $\tilde \beta$ parameter
estimation and the expected mean 
$\Delta \chi^2$ distribution over the sky,
respectively.
From both the plots, it is clear that the LSST may allow for a
significant detection of the aberration effect: the accuracy of the 
$\tilde \beta$ 
parameter estimation is at a level of $\sigma(\tilde \beta)=10^{-4}$, while the
dipole directions can be determined with the precision of $20$ degrees
in radius. 
If we can have a full-sky survey with similar depth to the
LSST, the ultimate precision is 10 degrees in radius.

\section{Summary and discussion}
\label{sec:Summary_Discussion}

In this paper we have used the SDSS DR6 galaxy catalog to explore the
dipole anisotropy pattern in the angular galaxy distribution caused by
the aberration effect due to the Earth's peculiar motion to the 
matter rest-frame.  The SDSS DR6 catalog, which covers about 8000 square
degrees, allows us to construct 
the 
currently most reliable galaxy catalog
for our purpose because the data has well-calibrated, homogeneous and
secure photometric and astrometric properties.

After developing a method to explore the aberration effect signal (see
Sec.~\ref{sec:Methodology}), we estimated the detectability of the
dipole signal for SDSS- and LSST-type surveys assuming the CMB dipole
amplitude and the concordance $\Lambda$CDM model properly taking into
account the contamination effects (Sec.~\ref{sec:rough_estimate}). We
found that the intrinsic galaxy clustering at low redshifts gives a
significant contamination.  Recalling that for a CDM model the galaxy
distribution has greater inhomogeneities at smaller scales, the
small-scale galaxy distribution at lower redshifts is viewed by an
observer with larger angles, and the strong inhomogeneities
significantly obscure the aberration effect if the Earth's peculiar
motion to the matter rest-frame is similar to that for the CMB dipole
amplitude, i.e. $\tilde \beta=O(10^{-3})$. In other words, the dipole
anisotropy due to the intrinsic clustering can be greater than the
aberration effect for a low-redshift galaxy catalog.  The Poisson
contamination due to the discrete galaxy distribution is smaller, but
not negligible even with such a huge number of SDSS galaxies. Therefore
our results imply that the previous report on a possible detection of
the aberration effect may be due to these contaminating effects
(see also \cite{Crawford09}).

To remove the contaminating effect of the low-redshift galaxies as much
as possible,  we used the
photometric redshift information to define a secure
catalog of galaxies. 
We considered the four galaxy catalogs to study 
possible remaining systematic effects due to the star-galaxy separation
and the intrinsic clustering contamination, as are defined in Table
\ref{tab:samp}.  As shown in Table~\ref{tbl:SDSS_chi2result} and
Fig.~\ref{fig:deltachi2_map}, the ``faint'' samples, which are supposed to
be more contaminated by an imperfect star-galaxy separation than for the
``bright'' samples, tend to prefer the dipole directions at higher
galactic latitudes. This is counter-intuitive, because the faint samples
may suffer from more star contamination 
towards lower galactic latitudes, which in turn causes 
the dipole direction in the lower
latitudes due to the enhanced overdensity. Hence the imperfect star-galaxy
separation seems not a main source of the systematic effects. We also
compared the ``deep'' and ``shallow'' samples by using the photometric
redshift information. As described above, the shallower sample is
supposed to be more affected by the intrinsic galaxy clustering
contamination from more evolving nonlinear structures at lower
redshifts. However, the dipole directions for both the 
deep and shallow samples are found to be similar, which indeed prefers
a direction around the CMB dipole direction within the $1$-$\sigma$
region, except for the 
``faint-deep'' sample (see Fig.~\ref{fig:deltachi2_map}). 

Among the 
four 
galaxy catalogs, the ``faint-deep'' and
``faint-shallow'' 
samples are found to give a possible detection of dipole
amplitude given as 
$\tilde \beta\sim 10^{-2}$
corresponding to the peculiar
velocity $v\sim 3000$km~s$^{-1}$ (see Fig.~\ref{fig:prob_beta}). These
values are significantly larger than the CMB dipole amplitude by a
factor 10. The peculiar velocity amplitude of $10^3$km~s$^{-1}$ is
similar to the virial velocities of a massive cluster, however, we know
our Galaxy is not in such a cluster. On the other hand, the large-scale
(relative) bulk flow of $1000$km~s$^{-1}$ is at $2-3\sigma$ deviations
from  the typical peculiar velocity 
of $470$km~s$^{-1}$ 
predicted by the standard $\Lambda$CDM structure formation
model. Furthermore, the previous studies on the peculiar velocity field
for the Local Group have not found such a large peculiar
velocity.
Therefore we believe that a possible indication on the
aberration effect on the SDSS galaxy distribution is not significant,
and may be due to 
residual contaminating effects.

We argued 
that the dipole direction may be caused, at least partly, 
by the apparent systematic 
difference of galaxy number densities between the SDSS survey regions of
the Northern and Southern Galactic hemispheres (NGH and SGH, respectively),
where the SGH region has a smaller average number density than the
NGH (see Table~\ref{tab:samp}). 
We found that the number density differences between NGH and SGH
does move the direction of the detected dipoles in all the four samples toward
the CMB dipole direction, however, the origin of the difference itself is
unknown.
The two regions are separated, and the SGH 
region has a much smaller sky coverage (679 deg$^2$) compared with the
NGH region (6928 deg$^2$). 
A data having a wider sky coverage is desired to settle if this
difference is real or due to some systematics in our analysis.

Another concern is on the Galactic dust extinction correction. 
Indeed, when excluding the SGH data, the preferred dipole directions of all
samples more or less move toward the Northern Galactic pole. Hence we
suspect that 
Galactic dust may cause one part of our detection of the dipoles in
the three less robust samples other than the BD sample (another possible part is 
the number density difference between NGH and SGH).  
Although we have assumed in this paper that the SFD dust map 
\cite{Schlegel1998} gives us 
a perfect dust extinction correction, Yahata et al. \cite{Yahata2007}
pointed our remaining systematics in low dust extinction regions  
in the map. 
Therefore an indication of the dipole anisotropy we found is not yet
conclusive.

We quantitatively showed that a survey with almost full-sky coverage and
sufficient depth
is ideally needed to explore the aberration
effect on the galaxy distribution. Future surveys, 
Pan-Starrs \citep[][see also http://pan-starrs.ifa.hawaii.edu/public/]{PanStarrs} or LSST \cite{LSST},
are such a survey and offer a chance to explore the signal. In this
paper we showed that LSST may allow for a significant detection of the
aberration effect. Even for the effect similar to the CMB dipole
amplitude, the angular direction of the aberration effect can be
determined with precision of 20 degrees in radius. 
If the cosmic bulk flow of order $\sim 1000$km~s$^{-1}$ implied 
in Kashlinsky et
al. (2008)\cite{Kashlinsky2008ApJL} extends out to the horizon and
therefore we are at rest 
in the matter rest-frame, the LSST survey will provide us a good opportunity for either 
accepting or rejecting such a bulk flow.   
Thus LSST may offer a
unique chance to constrain the horizon-scale perturbations where some
exotic physics related to cosmic acceleration such as dark energy and
super-void may play a role.

There are other methods/data-sets that allow to explore the horizon-scale
peculiar velocity field relative to the Earth's motion. 
For instance,
using a
homogeneous catalog of quasars, such as that of SDSS, is advantageous to
reduce the intrinsic clustering contamination due to the higher redshift
coverage. However a clean identification of quasars from stellar
population from imaging data is always problematic. Also the Poisson
noise can be significant due to a much smaller number density of
quasars.

Another interesting
method is using the 
kinetic Sunyaev-Zel'dovich (SZ) effect that is caused by the peculiar
velocities of ionized medium that scatters off CMB photons causing the
secondary temperature fluctuations by the Doppler shift. For example,
the kinetic SZ effect can be extracted by measuring the CMB
fluctuations. The advantage of this method is it allows to directly measure the
line-of-sight component of peculiar velocity at the cluster redshift.
In fact there are several attempts to measure the kinetic SZ effect by
stacking the CMB fluctuations in the sky regions of clusters in order to
explore the excess in the temperature fluctuations
\cite{Chluba2005,Kashlinsky2008ApJL,2009arXiv0903.2845H}. In
particular, 
Kashlinsky et al. (2008)\cite{Kashlinsky2008ApJL} reported a possible significant detection of
the large-scale bulk flow of $X$-ray luminous clusters, 
implying the bulk flow of $\sim 1000$km~s$^{-1}$, 
apparently similar to that
indicated from our analysis on our faint samples within the uncertainties, 
in the angular direction not far from the
 CMB dipole direction (also see
\cite{2009ApJ...691.1479K,Kashlinsky09}). 
Based on this surprising result, it was
speculated that the large-scale bulk flow may originate from the
non-standard horizon-scale perturbations, for example the tilt across
the observable Universe due to the pre-inflationary inhomogeneities
\cite{Turner1991}. This result is, very interesting though,
still under debate, and a further analysis will be needed by using a
homogeneous massive sample of clusters. Moreover, a
high-angular-resolution and high-sensitive CMB measurement has a
potential to extract the kinetic SZ effect due to the ionized
intergalactic medium (more exactly known as the Ostriker-Vishniac effect
\cite{1986ApJ...306L..51O}).
Thus it would be worth exploring
large-scale bulk flow of the Universe by combining various methods in
order to explore the possible new horizon-scale physics, which is very
difficult to explore by other means.

\acknowledgments

We would like to thank Masashi Chiba, Yuji Chinone, Makoto Hattori,
Nobuhiro Okabe, Tsutomu T. Takeuchi, An$\check{\rm z}$e Slosar, and 
Saleem Zaroubi for useful comments that 
clarifies statistical issues in our analysis. 
We are grateful to 
Ryuichi Takahashi and Guilhem Lavaux who let us know a couple of
interesting references.   
We thank the anonymous referee who carefully read our manuscript and gave
us interesting comments. We have used the FFTLog code 
(\verb+http://casa.colorado.edu/~ajsh/FFTLog/+) written by professor Andrew Hamilton 
to whom we are grateful. K. Y. was supported by Grants-in-Aid for Japan Society for 
the Promotion of Science Fellows. 
Y. I. was supported by a Grant-in-Aid for the 21st Century Center of
Excellence (COE)
Program ``Exploring New Science by Bridging Particle-Matter Hierarchy'' 
and is supported by the Japan Society of the
Promotion of Science Global COE Program (G01): 
``Weaving Science Web beyond Particle-Matter Hierarchy'', both 
in Tohoku University, and both funded by the Ministry of Education, 
Science, Sports and Culture of Japan. 
Funding for the SDSS and SDSS-II has been provided by the Alfred
P. Sloan Foundation, 
the Participating Institutions, the National Science Foundation, the
U.S. Department of Energy, 
the National Aeronautics and Space Administration, the Japanese
Monbukagakusho, the Max Planck Society, 
and the Higher Education Funding Council for England. The SDSS Web Site is http://www.sdss.org/.
The SDSS is managed by the Astrophysical Research Consortium for the
Participating Institutions. 
The Participating Institutions are the American Museum of Natural
History, Astrophysical Institute Potsdam, 
University of Basel, University of Cambridge, Case Western Reserve
University, University of Chicago, 
Drexel University, Fermilab, the Institute for Advanced Study, the Japan
Participation Group, 
Johns Hopkins University, the Joint Institute for Nuclear Astrophysics, 
the Kavli Institute for Particle Astrophysics and Cosmology, the Korean
Scientist Group, 
the Chinese Academy of Sciences (LAMOST), Los Alamos National
Laboratory, 
the Max-Planck-Institute for Astronomy (MPIA), the Max-Planck-Institute
for Astrophysics (MPA), 
New Mexico State University, Ohio State University, University of
Pittsburgh, University of Portsmouth, 
Princeton University, the United States Naval Observatory, and the University of Washington.

\appendix
\section{Dipole Amplitudes}
\label{sec:appendix:Dipole_Amplitudes}

This appendix briefly explains a method of estimating 
the amplitudes of the expected dipole moments in the galaxy 
density fluctuation field from three contributions: the Poisson shot noise, 
the large scale structure, and the motion of us (the Earth) with 
respect to the CMB rest frame. This method is used in 
Sec.~\ref{sec:rough_estimate} to obtain a rough estimate of a signal to
noise ratio for detecting a dipole modulation.  We first consider   
the case of a hypothetical all-sky survey, then a 
partial-sky survey.  We follow the method of  
\cite{Peebles1973,Scharf1992,Baleisis1998} with which readers' may
consult for further details.

\subsection{All-Sky Survey}
This section considers a hypothetical all-sky survey  
and we will start with our definitions of the dipole moments. 

For a galaxy number density field $n(\bi{\theta})$, 
we compute a spherical harmonic expansion of $n(\bi{\theta})$ as 
\begin{eqnarray}
a_l^m = \int_{\Omega_S}d\Omega_{\bi{\theta}} n(\bi{\theta})Y_l^{m*}(\bi{\theta}),
\label{eq:def_alm}
\end{eqnarray}
where the integral is over the survey area $\Omega_S$ and 
$^*$ denotes a complex conjugate.
We shall use the convention of \cite{Jackson} for the spherical
harmonics $Y_l^m(\bi{\theta})$. The dipole vector components in the three dimensional 
Cartesian coordinates is then defined as 
$
\vec D
=
(
-\sqrt{2}{\cal R}e(a_{1}^1),
\sqrt{2}{\cal I}m(a_{1}^1),
a_{1}^0
) 
$ where ${\cal R}e(f)$ and ${\cal I}m(f)$ stand for a real and an
imaginary part of some quantity $f$, respectively (See, e.g., \cite{Copi2004}). 
Accordingly, the magnitude of the dipole is 
\begin{eqnarray}
\langle|\vec D|^2\rangle = 
\sum_{m=-1}^1\left\langle|a_1^{m}|^2\right\rangle,
\label{eq:def_dipole_modulus}
\end{eqnarray}
where the angle brackets denote an ensemble average.
Given $a_l^m$, we can estimate the magnitude of the dipole moments of 
various contributions.

\subsubsection{Aberration Induced Dipole}

If the aberration is the only cause of the dipole in the galaxy 
number density field, the field has the form given by
Eq. (\ref{eqn:del-rel}). 
We define the dipole amplitude 
due to the aberration effect $A_{\ab}$ as 
\begin{eqnarray}
\sqrt{|\vec D_{\ab}|^2} &\equiv& \sqrt{\frac{4\pi}{3}}\bar n_g A_{\ab}.
\label{eq:aberration_dipole_amplitude}
\end{eqnarray} 
where $\bar n_g$ is the average angular number density of galaxies per steradian. 
By evaluating Eqs. (\ref{eq:def_alm}) and (\ref{eq:def_dipole_modulus})
using Eq. (\ref{eqn:del-rel}), 
we can find $A_{\ab} = 2\tilde \beta$ for an all-sky survey (and this is
the reason for factoring out $\bar n_g \sqrt{4\pi/3}$ in the right hand side of the
above equation).

\subsubsection{Large Scale Structure Dipole}

The coefficient $a_l^m$ from the large scale structure (LSS) 
may be estimated as 
\begin{eqnarray}
\langle|a_{l,\LSS}^m|^2\rangle = 
\frac{2}{\pi}
\bar n_g^2 
b_g^2\int^{\infty}_0P_m^L(k)|I_l(k)|^2k^2dk
= \bar n_g^2 C_g(l).
\end{eqnarray}
We then define the amplitude of the dipole due to large scale structure
$A_{\LSS}$ as   
\begin{eqnarray}
\sqrt{\langle|\vec D_{\LSS}|^2\rangle} &=& 
 \sqrt{3 C_g(1)}\bar n_g
\equiv \sqrt{\frac{4\pi}{3}}\bar n_g A_{\LSS}.
\label{eq:LSS_dipole_amplitude}
\end{eqnarray}

\subsubsection{Poisson Shot Noise Dipole}

For the Poisson shot noise, $\langle a_{l,{\rm P}}^m\rangle = \bar n_g$. 
We define the amplitude of the dipole due to Poisson noise $A_{\rm P}$ as 
\begin{eqnarray}
\sqrt{\langle|\vec D_{\rm P}|^2\rangle} &=& 
 \sqrt{3 \bar n_g} 
\equiv \sqrt{\frac{4\pi}{3}}\bar n_g A_{\rm P}.
\label{eq:Poisson_dipole_amplitude}
\end{eqnarray}

\subsection{Partial-Sky Survey}

The incomplete sky coverage may be taken into account by 
a tensor $W_{ll'}^{mm'}$ \citep{Peebles1973,Scharf1992}.
It is defined as 
\begin{eqnarray}
W_{ll'}^{mm'} &=& \oint_{4\pi}d\Omega_{\bi{\theta}} Y_l^m(\bi{\theta})Y_{l'}^{m'*}(\bi{\theta})M(\Omega),
\end{eqnarray} 
where an angular 'mask' $M(\Omega)$ is defined such that $M(\Omega) = 1$ 
over the area of the sky observed and $M(\Omega) = 0$ elsewhere.
This tensor satisfies for any $l$ 
\begin{eqnarray}
\frac{1}{2l+1}\sum_{m=-l}^l\sum_{l'=0}^{\infty}\sum_{m'=-l'}^{l'}
|W_{ll'}^{mm'}|^2 &=& 
\frac{\Omega_{S}}{4\pi}.
\label{eq:W_tensor_properties1}
\end{eqnarray} 

With the tensor $W_{ll'}^{mm'}$, the 
coefficient of the spherical harmonic expansion 
for the incomplete sky coverage becomes 
\begin{eqnarray}
\langle|c_l^m|^2\rangle &=& \sum_{l'=0}^{\infty}
\sum_{m'=^-l'}^{l'}|W_{ll'}^{mm'}|^2\langle|a_{l'}^{m'}|^2\rangle, 
\end{eqnarray} 
where $a_l^m$ is the coefficients for the all-sky survey.
We then define the amplitudes of the dipole fluctuation in the 
galaxy number density field in terms of $c_l^m$ instead of 
$a_l^m$. Introducing coefficients $S_{ll'}$ as 
\begin{eqnarray}
S_{ll'} &=&
\sum_{m=-l}^l\sum_{m'=-l'}^{l'}|W_{ll'}^{mm'}|^2,
\end{eqnarray} 
we define the dipole amplitudes 
from various contributions 
for a partial-sky survey as  
\begin{eqnarray}
A_{LSS} &=& 
\sqrt{\frac{3}{4\pi}\sum_{l' = 0}^{\infty}S_{1l'} C_g({l'})},
\label{eq:LSS_dipole_amplitude_partial_sky}
\\
A_{\rm P} &=& 
\frac{3}{\sqrt{4\pi}}
\sqrt{ \frac{\Omega_{S}}{4\pi \bar n_g}}, 
\label{eq:Poisson_dipole_amplitude_partial_sky}
\end{eqnarray} 
and 
\begin{eqnarray}
A_{\ab} &=& 
2\tilde \beta 
\sqrt{\frac{S_{11'}}{3}}, 
\label{eq:aberration_dipole_amplitude_partial_sky}
\end{eqnarray}
which are the equations used to obtain the rough estimates of the signal to
noise ratio of the aberration induced dipole in Sec.~\ref{sec:rough_estimate}.

Finally, we point out two things. Firstly, for a full-sky survey,
$W_{ll'}^{mm'}=\delta^K_{ll'}\delta^K_{mm' }$ and hence  
$S_{ll'} = (2l+1)\delta^K_{ll'}$. As a result, we recover $A_{\rm ab} =
2\tilde \beta$ for such a survey as is expected.
Secondly, we note that in a practical computation, we cannot complete the $l'$ 
summation in Eq. (\ref{eq:LSS_dipole_amplitude_partial_sky}). 
We have made the summation up to $l' = 500$ for the SDSS DR6 
survey geometry and $l'=200$ for the expected LSST survey 
geometry considered in this paper. Indeed, it is found that 
Eq. (\ref{eq:W_tensor_properties1}) satisfies to a good accuracy 
for these upper limit values for the $l'$-summation. 
\begin{eqnarray}
\frac{1}{3}\sum_{m=-1}^{m=1}\sum_{l'=0}^{l'=500}\sum_{m'=-l'}^{m=l'}
|W_{1l'}^{mm'}|^2 &=& 
0.9455\frac{\Omega_{SDSS}}{4\pi},
\nonumber \\
\frac{1}{3}\sum_{m=-1}^{m=1}\sum_{l'=0}^{l'=200}\sum_{m'=-l'}^{m=l'}
|W_{1l'}^{mm'}|^2 &=& 
0.9966\frac{\Omega_{LSST}}{4\pi}.
\nonumber 
\end{eqnarray}

%\bibliography{refs_091207}

\begin{thebibliography}{58}
\expandafter\ifx\csname natexlab\endcsname\relax\def\natexlab#1{#1}\fi
\expandafter\ifx\csname bibnamefont\endcsname\relax
  \def\bibnamefont#1{#1}\fi
\expandafter\ifx\csname bibfnamefont\endcsname\relax
  \def\bibfnamefont#1{#1}\fi
\expandafter\ifx\csname citenamefont\endcsname\relax
  \def\citenamefont#1{#1}\fi
\expandafter\ifx\csname url\endcsname\relax
  \def\url#1{\texttt{#1}}\fi
\expandafter\ifx\csname urlprefix\endcsname\relax\def\urlprefix{URL }\fi
\providecommand{\bibinfo}[2]{#2}
\providecommand{\eprint}[2][]{\url{#2}}

\bibitem[{\citenamefont{{Kogut} et~al.}(1993)\citenamefont{{Kogut},
  {Lineweaver}, {Smoot}, {Bennett}, {Banday}, {Boggess}, {Cheng}, {de Amici},
  {Fixsen}, {Hinshaw} et~al.}}]{Kogut1993}
\bibinfo{author}{\bibfnamefont{A.}~\bibnamefont{{Kogut}}},
  \bibinfo{author}{\bibfnamefont{C.}~\bibnamefont{{Lineweaver}}},
  \bibinfo{author}{\bibfnamefont{G.~F.} \bibnamefont{{Smoot}}},
  \bibinfo{author}{\bibfnamefont{C.~L.} \bibnamefont{{Bennett}}},
  \bibinfo{author}{\bibfnamefont{A.}~\bibnamefont{{Banday}}},
  \bibinfo{author}{\bibfnamefont{N.~W.} \bibnamefont{{Boggess}}},
  \bibinfo{author}{\bibfnamefont{E.~S.} \bibnamefont{{Cheng}}},
  \bibinfo{author}{\bibfnamefont{G.}~\bibnamefont{{de Amici}}},
  \bibinfo{author}{\bibfnamefont{D.~J.} \bibnamefont{{Fixsen}}},
  \bibinfo{author}{\bibfnamefont{G.}~\bibnamefont{{Hinshaw}}},
  \bibnamefont{et~al.}, \bibinfo{journal}{\apj} \textbf{\bibinfo{volume}{419}},
  \bibinfo{pages}{1} (\bibinfo{year}{1993}).

\bibitem[{\citenamefont{{Lineweaver} et~al.}(1996)\citenamefont{{Lineweaver},
  {Tenorio}, {Smoot}, {Keegstra}, {Banday}, and {Lubin}}}]{Lineweaver1996}
\bibinfo{author}{\bibfnamefont{C.~H.} \bibnamefont{{Lineweaver}}},
  \bibinfo{author}{\bibfnamefont{L.}~\bibnamefont{{Tenorio}}},
  \bibinfo{author}{\bibfnamefont{G.~F.} \bibnamefont{{Smoot}}},
  \bibinfo{author}{\bibfnamefont{P.}~\bibnamefont{{Keegstra}}},
  \bibinfo{author}{\bibfnamefont{A.~J.} \bibnamefont{{Banday}}},
  \bibnamefont{and} \bibinfo{author}{\bibfnamefont{P.}~\bibnamefont{{Lubin}}},
  \bibinfo{journal}{\apj} \textbf{\bibinfo{volume}{470}}, \bibinfo{pages}{38}
  (\bibinfo{year}{1996}).

\bibitem[{\citenamefont{{Courteau} and {van den Bergh}}(1999)}]{Courteau1999}
\bibinfo{author}{\bibfnamefont{S.}~\bibnamefont{{Courteau}}} \bibnamefont{and}
  \bibinfo{author}{\bibfnamefont{S.}~\bibnamefont{{van den Bergh}}},
  \bibinfo{journal}{\aj} \textbf{\bibinfo{volume}{118}}, \bibinfo{pages}{337}
  (\bibinfo{year}{1999}).

\bibitem[{\citenamefont{{Dehnen} and {Binney}}(1998)}]{DehnenBinney1998}
\bibinfo{author}{\bibfnamefont{W.}~\bibnamefont{{Dehnen}}} \bibnamefont{and}
  \bibinfo{author}{\bibfnamefont{J.~J.} \bibnamefont{{Binney}}},
  \bibinfo{journal}{\mnras} \textbf{\bibinfo{volume}{298}},
  \bibinfo{pages}{387} (\bibinfo{year}{1998}).

\bibitem[{\citenamefont{{Reid} et~al.}(2009)\citenamefont{{Reid}, {Menten},
  {Zheng}, {Brunthaler}, {Moscadelli}, {Xu}, {Zhang}, {Sato}, {Honma}, {Hirota}
  et~al.}}]{Reid2009}
\bibinfo{author}{\bibfnamefont{M.~J.} \bibnamefont{{Reid}}},
  \bibinfo{author}{\bibfnamefont{K.~M.} \bibnamefont{{Menten}}},
  \bibinfo{author}{\bibfnamefont{X.~W.} \bibnamefont{{Zheng}}},
  \bibinfo{author}{\bibfnamefont{A.}~\bibnamefont{{Brunthaler}}},
  \bibinfo{author}{\bibfnamefont{L.}~\bibnamefont{{Moscadelli}}},
  \bibinfo{author}{\bibfnamefont{Y.}~\bibnamefont{{Xu}}},
  \bibinfo{author}{\bibfnamefont{B.}~\bibnamefont{{Zhang}}},
  \bibinfo{author}{\bibfnamefont{M.}~\bibnamefont{{Sato}}},
  \bibinfo{author}{\bibfnamefont{M.}~\bibnamefont{{Honma}}},
  \bibinfo{author}{\bibfnamefont{T.}~\bibnamefont{{Hirota}}},
  \bibnamefont{et~al.}, \bibinfo{journal}{\apj} \textbf{\bibinfo{volume}{700}},
  \bibinfo{pages}{137} (\bibinfo{year}{2009}).

\bibitem[{\citenamefont{{Lynden-Bell} et~al.}(1989)\citenamefont{{Lynden-Bell},
  {Lahav}, and {Burstein}}}]{Lynden-Belletal89}
\bibinfo{author}{\bibfnamefont{D.}~\bibnamefont{{Lynden-Bell}}},
  \bibinfo{author}{\bibfnamefont{O.}~\bibnamefont{{Lahav}}}, \bibnamefont{and}
  \bibinfo{author}{\bibfnamefont{D.}~\bibnamefont{{Burstein}}},
  \bibinfo{journal}{\mnras} \textbf{\bibinfo{volume}{241}},
  \bibinfo{pages}{325} (\bibinfo{year}{1989}).

\bibitem[{\citenamefont{{Strauss} et~al.}(1992)\citenamefont{{Strauss},
  {Yahil}, {Davis}, {Huchra}, and {Fisher}}}]{Straussetal92}
\bibinfo{author}{\bibfnamefont{M.~A.} \bibnamefont{{Strauss}}},
  \bibinfo{author}{\bibfnamefont{A.}~\bibnamefont{{Yahil}}},
  \bibinfo{author}{\bibfnamefont{M.}~\bibnamefont{{Davis}}},
  \bibinfo{author}{\bibfnamefont{J.~P.} \bibnamefont{{Huchra}}},
  \bibnamefont{and} \bibinfo{author}{\bibfnamefont{K.}~\bibnamefont{{Fisher}}},
  \bibinfo{journal}{\apj} \textbf{\bibinfo{volume}{397}}, \bibinfo{pages}{395}
  (\bibinfo{year}{1992}).

\bibitem[{\citenamefont{{Zaroubi}}(2002)}]{Saleem}
\bibinfo{author}{\bibfnamefont{S.}~\bibnamefont{{Zaroubi}}},
  \bibinfo{journal}{ArXiv Astrophysics e-prints}  (\bibinfo{year}{2002}),
  \eprint{arXiv:astro-ph/0206052}.

\bibitem[{\citenamefont{{Erdo{\u g}du} and {Lahav}}(2009)}]{Erdogdu2009}
\bibinfo{author}{\bibfnamefont{P.}~\bibnamefont{{Erdo{\u g}du}}}
  \bibnamefont{and} \bibinfo{author}{\bibfnamefont{O.}~\bibnamefont{{Lahav}}},
  \bibinfo{journal}{\prd} \textbf{\bibinfo{volume}{80}},
  \bibinfo{pages}{043005} (\bibinfo{year}{2009}).

\bibitem[{\citenamefont{{Basilakos} and
  {Plionis}}(2006)}]{BasilakosPlionis2006}
\bibinfo{author}{\bibfnamefont{S.}~\bibnamefont{{Basilakos}}} \bibnamefont{and}
  \bibinfo{author}{\bibfnamefont{M.}~\bibnamefont{{Plionis}}},
  \bibinfo{journal}{\mnras} \textbf{\bibinfo{volume}{373}},
  \bibinfo{pages}{1112} (\bibinfo{year}{2006}).

\bibitem[{\citenamefont{{Kocevski} and {Ebeling}}(2006)}]{KocevskiEbeling2006}
\bibinfo{author}{\bibfnamefont{D.~D.} \bibnamefont{{Kocevski}}}
  \bibnamefont{and}
  \bibinfo{author}{\bibfnamefont{H.}~\bibnamefont{{Ebeling}}},
  \bibinfo{journal}{\apj} \textbf{\bibinfo{volume}{645}}, \bibinfo{pages}{1043}
  (\bibinfo{year}{2006}).

\bibitem[{\citenamefont{{Erdo{\u g}du} et~al.}(2006)\citenamefont{{Erdo{\u
  g}du}, {Huchra}, {Lahav}, {Colless}, {Cutri}, {Falco}, {George}, {Jarrett},
  {Jones}, {Kochanek} et~al.}}]{Erdogdu2006}
\bibinfo{author}{\bibfnamefont{P.}~\bibnamefont{{Erdo{\u g}du}}},
  \bibinfo{author}{\bibfnamefont{J.~P.} \bibnamefont{{Huchra}}},
  \bibinfo{author}{\bibfnamefont{O.}~\bibnamefont{{Lahav}}},
  \bibinfo{author}{\bibfnamefont{M.}~\bibnamefont{{Colless}}},
  \bibinfo{author}{\bibfnamefont{R.~M.} \bibnamefont{{Cutri}}},
  \bibinfo{author}{\bibfnamefont{E.}~\bibnamefont{{Falco}}},
  \bibinfo{author}{\bibfnamefont{T.}~\bibnamefont{{George}}},
  \bibinfo{author}{\bibfnamefont{T.}~\bibnamefont{{Jarrett}}},
  \bibinfo{author}{\bibfnamefont{D.~H.} \bibnamefont{{Jones}}},
  \bibinfo{author}{\bibfnamefont{C.~S.} \bibnamefont{{Kochanek}}},
  \bibnamefont{et~al.}, \bibinfo{journal}{\mnras}
  \textbf{\bibinfo{volume}{368}}, \bibinfo{pages}{1515} (\bibinfo{year}{2006}).

\bibitem[{\citenamefont{{Watkins} et~al.}(2009)\citenamefont{{Watkins},
  {Feldman}, and {Hudson}}}]{Watkins2009}
\bibinfo{author}{\bibfnamefont{R.}~\bibnamefont{{Watkins}}},
  \bibinfo{author}{\bibfnamefont{H.~A.} \bibnamefont{{Feldman}}},
  \bibnamefont{and} \bibinfo{author}{\bibfnamefont{M.~J.}
  \bibnamefont{{Hudson}}}, \bibinfo{journal}{\mnras}
  \textbf{\bibinfo{volume}{392}}, \bibinfo{pages}{743} (\bibinfo{year}{2009}).

\bibitem[{\citenamefont{{Lavaux} et~al.}(2010)\citenamefont{{Lavaux}, {Tully},
  {Mohayaee}, and {Colombi}}}]{Lavaux2010}
\bibinfo{author}{\bibfnamefont{G.}~\bibnamefont{{Lavaux}}},
  \bibinfo{author}{\bibfnamefont{R.~B.} \bibnamefont{{Tully}}},
  \bibinfo{author}{\bibfnamefont{R.}~\bibnamefont{{Mohayaee}}},
  \bibnamefont{and}
  \bibinfo{author}{\bibfnamefont{S.}~\bibnamefont{{Colombi}}},
  \bibinfo{journal}{\apj} \textbf{\bibinfo{volume}{709}}, \bibinfo{pages}{483}
  (\bibinfo{year}{2010}).

\bibitem[{\citenamefont{{Ellis} and {Baldwin}}(1984)}]{EllisBaldwin1984}
\bibinfo{author}{\bibfnamefont{G.~F.~R.} \bibnamefont{{Ellis}}}
  \bibnamefont{and} \bibinfo{author}{\bibfnamefont{J.~E.}
  \bibnamefont{{Baldwin}}}, \bibinfo{journal}{\mnras}
  \textbf{\bibinfo{volume}{206}}, \bibinfo{pages}{377} (\bibinfo{year}{1984}).

\bibitem[{\citenamefont{{Turner}}(1991)}]{Turner1991}
\bibinfo{author}{\bibfnamefont{M.~S.} \bibnamefont{{Turner}}},
  \bibinfo{journal}{\prd} \textbf{\bibinfo{volume}{44}}, \bibinfo{pages}{3737}
  (\bibinfo{year}{1991}).

\bibitem[{\citenamefont{{Caldwell} et~al.}(1998)\citenamefont{{Caldwell},
  {Dave}, and {Steinhardt}}}]{Caldwelletal98}
\bibinfo{author}{\bibfnamefont{R.~R.} \bibnamefont{{Caldwell}}},
  \bibinfo{author}{\bibfnamefont{R.}~\bibnamefont{{Dave}}}, \bibnamefont{and}
  \bibinfo{author}{\bibfnamefont{P.~J.} \bibnamefont{{Steinhardt}}},
  \bibinfo{journal}{Physical Review Letters} \textbf{\bibinfo{volume}{80}},
  \bibinfo{pages}{1582} (\bibinfo{year}{1998}).

\bibitem[{\citenamefont{{Takada}}(2006)}]{Takada06}
\bibinfo{author}{\bibfnamefont{M.}~\bibnamefont{{Takada}}},
  \bibinfo{journal}{\prd} \textbf{\bibinfo{volume}{74}},
  \bibinfo{pages}{043505} (\bibinfo{year}{2006}).

\bibitem[{\citenamefont{{Bennett} et~al.}(1996)\citenamefont{{Bennett},
  {Banday}, {Gorski}, {Hinshaw}, {Jackson}, {Keegstra}, {Kogut}, {Smoot},
  {Wilkinson}, and {Wright}}}]{Bennettetal96}
\bibinfo{author}{\bibfnamefont{C.~L.} \bibnamefont{{Bennett}}},
  \bibinfo{author}{\bibfnamefont{A.~J.} \bibnamefont{{Banday}}},
  \bibinfo{author}{\bibfnamefont{K.~M.} \bibnamefont{{Gorski}}},
  \bibinfo{author}{\bibfnamefont{G.}~\bibnamefont{{Hinshaw}}},
  \bibinfo{author}{\bibfnamefont{P.}~\bibnamefont{{Jackson}}},
  \bibinfo{author}{\bibfnamefont{P.}~\bibnamefont{{Keegstra}}},
  \bibinfo{author}{\bibfnamefont{A.}~\bibnamefont{{Kogut}}},
  \bibinfo{author}{\bibfnamefont{G.~F.} \bibnamefont{{Smoot}}},
  \bibinfo{author}{\bibfnamefont{D.~T.} \bibnamefont{{Wilkinson}}},
  \bibnamefont{and} \bibinfo{author}{\bibfnamefont{E.~L.}
  \bibnamefont{{Wright}}}, \bibinfo{journal}{\apjl}
  \textbf{\bibinfo{volume}{464}}, \bibinfo{pages}{L1} (\bibinfo{year}{1996}).

\bibitem[{\citenamefont{{Jing} and {Fang}}(1994)}]{Jingetal94}
\bibinfo{author}{\bibfnamefont{Y.}~\bibnamefont{{Jing}}} \bibnamefont{and}
  \bibinfo{author}{\bibfnamefont{L.}~\bibnamefont{{Fang}}},
  \bibinfo{journal}{Physical Review Letters} \textbf{\bibinfo{volume}{73}},
  \bibinfo{pages}{1882} (\bibinfo{year}{1994}).

\bibitem[{\citenamefont{{Efstathiou}}(2003)}]{Efstathiou03}
\bibinfo{author}{\bibfnamefont{G.}~\bibnamefont{{Efstathiou}}},
  \bibinfo{journal}{\mnras} \textbf{\bibinfo{volume}{343}},
  \bibinfo{pages}{L95} (\bibinfo{year}{2003}).

\bibitem[{\citenamefont{{Baleisis}
  et~al.}(1998{\natexlab{a}})\citenamefont{{Baleisis}, {Lahav}, {Loan}, and
  {Wall}}}]{Baleisisetal98}
\bibinfo{author}{\bibfnamefont{A.}~\bibnamefont{{Baleisis}}},
  \bibinfo{author}{\bibfnamefont{O.}~\bibnamefont{{Lahav}}},
  \bibinfo{author}{\bibfnamefont{A.~J.} \bibnamefont{{Loan}}},
  \bibnamefont{and} \bibinfo{author}{\bibfnamefont{J.~V.}
  \bibnamefont{{Wall}}}, \bibinfo{journal}{\mnras}
  \textbf{\bibinfo{volume}{297}}, \bibinfo{pages}{545}
  (\bibinfo{year}{1998}{\natexlab{a}}).

\bibitem[{\citenamefont{{Scharf} et~al.}(2000)\citenamefont{{Scharf}, {Jahoda},
  {Treyer}, {Lahav}, {Boldt}, and {Piran}}}]{Scharfetal00}
\bibinfo{author}{\bibfnamefont{C.~A.} \bibnamefont{{Scharf}}},
  \bibinfo{author}{\bibfnamefont{K.}~\bibnamefont{{Jahoda}}},
  \bibinfo{author}{\bibfnamefont{M.}~\bibnamefont{{Treyer}}},
  \bibinfo{author}{\bibfnamefont{O.}~\bibnamefont{{Lahav}}},
  \bibinfo{author}{\bibfnamefont{E.}~\bibnamefont{{Boldt}}}, \bibnamefont{and}
  \bibinfo{author}{\bibfnamefont{T.}~\bibnamefont{{Piran}}},
  \bibinfo{journal}{\apj} \textbf{\bibinfo{volume}{544}}, \bibinfo{pages}{49}
  (\bibinfo{year}{2000}).

\bibitem[{\citenamefont{{Blake} and {Wall}}(2002)}]{BlakeWall2002}
\bibinfo{author}{\bibfnamefont{C.}~\bibnamefont{{Blake}}} \bibnamefont{and}
  \bibinfo{author}{\bibfnamefont{J.}~\bibnamefont{{Wall}}},
  \bibinfo{journal}{\nat} \textbf{\bibinfo{volume}{416}}, \bibinfo{pages}{150}
  (\bibinfo{year}{2002}).

\bibitem[{\citenamefont{{Condon} et~al.}(1998)\citenamefont{{Condon}, {Cotton},
  {Greisen}, {Yin}, {Perley}, {Taylor}, and {Broderick}}}]{NRAOVLASkySurvey}
\bibinfo{author}{\bibfnamefont{J.~J.} \bibnamefont{{Condon}}},
  \bibinfo{author}{\bibfnamefont{W.~D.} \bibnamefont{{Cotton}}},
  \bibinfo{author}{\bibfnamefont{E.~W.} \bibnamefont{{Greisen}}},
  \bibinfo{author}{\bibfnamefont{Q.~F.} \bibnamefont{{Yin}}},
  \bibinfo{author}{\bibfnamefont{R.~A.} \bibnamefont{{Perley}}},
  \bibinfo{author}{\bibfnamefont{G.~B.} \bibnamefont{{Taylor}}},
  \bibnamefont{and} \bibinfo{author}{\bibfnamefont{J.~J.}
  \bibnamefont{{Broderick}}}, \bibinfo{journal}{\aj}
  \textbf{\bibinfo{volume}{115}}, \bibinfo{pages}{1693} (\bibinfo{year}{1998}).

\bibitem[{\citenamefont{{York} et~al.}(2000)\citenamefont{{York}, {Adelman},
  {Anderson}, {Anderson}, {Annis}, {Bahcall}, {Bakken}, {Barkhouser},
  {Bastian}, {Berman} et~al.}}]{York2000}
\bibinfo{author}{\bibfnamefont{D.~G.} \bibnamefont{{York}}},
  \bibinfo{author}{\bibfnamefont{J.}~\bibnamefont{{Adelman}}},
  \bibinfo{author}{\bibfnamefont{J.~E.} \bibnamefont{{Anderson}},
  \bibfnamefont{Jr.}}, \bibinfo{author}{\bibfnamefont{S.~F.}
  \bibnamefont{{Anderson}}},
  \bibinfo{author}{\bibfnamefont{J.}~\bibnamefont{{Annis}}},
  \bibinfo{author}{\bibfnamefont{N.~A.} \bibnamefont{{Bahcall}}},
  \bibinfo{author}{\bibfnamefont{J.~A.} \bibnamefont{{Bakken}}},
  \bibinfo{author}{\bibfnamefont{R.}~\bibnamefont{{Barkhouser}}},
  \bibinfo{author}{\bibfnamefont{S.}~\bibnamefont{{Bastian}}},
  \bibinfo{author}{\bibfnamefont{E.}~\bibnamefont{{Berman}}},
  \bibnamefont{et~al.}, \bibinfo{journal}{\aj} \textbf{\bibinfo{volume}{120}},
  \bibinfo{pages}{1579} (\bibinfo{year}{2000}).

\bibitem[{\citenamefont{{Spergel} et~al.}(2007)\citenamefont{{Spergel}, {Bean},
  {Dor{\'e}}, {Nolta}, {Bennett}, {Dunkley}, {Hinshaw}, {Jarosik}, {Komatsu},
  {Page} et~al.}}]{Spergel2007}
\bibinfo{author}{\bibfnamefont{D.~N.} \bibnamefont{{Spergel}}},
  \bibinfo{author}{\bibfnamefont{R.}~\bibnamefont{{Bean}}},
  \bibinfo{author}{\bibfnamefont{O.}~\bibnamefont{{Dor{\'e}}}},
  \bibinfo{author}{\bibfnamefont{M.~R.} \bibnamefont{{Nolta}}},
  \bibinfo{author}{\bibfnamefont{C.~L.} \bibnamefont{{Bennett}}},
  \bibinfo{author}{\bibfnamefont{J.}~\bibnamefont{{Dunkley}}},
  \bibinfo{author}{\bibfnamefont{G.}~\bibnamefont{{Hinshaw}}},
  \bibinfo{author}{\bibfnamefont{N.}~\bibnamefont{{Jarosik}}},
  \bibinfo{author}{\bibfnamefont{E.}~\bibnamefont{{Komatsu}}},
  \bibinfo{author}{\bibfnamefont{L.}~\bibnamefont{{Page}}},
  \bibnamefont{et~al.}, \bibinfo{journal}{\apjs}
  \textbf{\bibinfo{volume}{170}}, \bibinfo{pages}{377} (\bibinfo{year}{2007}).

\bibitem[{\citenamefont{{Fukugita} et~al.}(2004)\citenamefont{{Fukugita},
  {Yasuda}, {Brinkmann}, {Gunn}, {Ivezi{\'c}}, {Knapp}, {Lupton}, and
  {Schneider}}}]{Fukugita2004}
\bibinfo{author}{\bibfnamefont{M.}~\bibnamefont{{Fukugita}}},
  \bibinfo{author}{\bibfnamefont{N.}~\bibnamefont{{Yasuda}}},
  \bibinfo{author}{\bibfnamefont{J.}~\bibnamefont{{Brinkmann}}},
  \bibinfo{author}{\bibfnamefont{J.~E.} \bibnamefont{{Gunn}}},
  \bibinfo{author}{\bibfnamefont{{\v Z}.}~\bibnamefont{{Ivezi{\'c}}}},
  \bibinfo{author}{\bibfnamefont{G.~R.} \bibnamefont{{Knapp}}},
  \bibinfo{author}{\bibfnamefont{R.}~\bibnamefont{{Lupton}}}, \bibnamefont{and}
  \bibinfo{author}{\bibfnamefont{D.~P.} \bibnamefont{{Schneider}}},
  \bibinfo{journal}{\aj} \textbf{\bibinfo{volume}{127}}, \bibinfo{pages}{3155}
  (\bibinfo{year}{2004}).

\bibitem[{\citenamefont{{Bruzual} and {Charlot}}(2003)}]{BruzualCharlot2003}
\bibinfo{author}{\bibfnamefont{G.}~\bibnamefont{{Bruzual}}} \bibnamefont{and}
  \bibinfo{author}{\bibfnamefont{S.}~\bibnamefont{{Charlot}}},
  \bibinfo{journal}{\mnras} \textbf{\bibinfo{volume}{344}},
  \bibinfo{pages}{1000} (\bibinfo{year}{2003}).

\bibitem[{\citenamefont{{Stabenau} et~al.}(2008)\citenamefont{{Stabenau},
  {Connolly}, and {Jain}}}]{Stabenau2008}
\bibinfo{author}{\bibfnamefont{H.~F.} \bibnamefont{{Stabenau}}},
  \bibinfo{author}{\bibfnamefont{A.}~\bibnamefont{{Connolly}}},
  \bibnamefont{and} \bibinfo{author}{\bibfnamefont{B.}~\bibnamefont{{Jain}}},
  \bibinfo{journal}{\mnras} \textbf{\bibinfo{volume}{387}},
  \bibinfo{pages}{1215} (\bibinfo{year}{2008}).

\bibitem[{\citenamefont{{Peebles}}(1980)}]{Peebles80}
\bibinfo{author}{\bibfnamefont{P.~J.~E.} \bibnamefont{{Peebles}}},
  \emph{\bibinfo{title}{{The large-scale structure of the universe}}}
  (\bibinfo{publisher}{Princeton, N.J., Princeton University Press},
  \bibinfo{year}{1980}).

\bibitem[{\citenamefont{{Takada} and {Bridle}}(2007)}]{TakadaBridle07}
\bibinfo{author}{\bibfnamefont{M.}~\bibnamefont{{Takada}}} \bibnamefont{and}
  \bibinfo{author}{\bibfnamefont{S.}~\bibnamefont{{Bridle}}},
  \bibinfo{journal}{New Journal of Physics} \textbf{\bibinfo{volume}{9}},
  \bibinfo{pages}{446} (\bibinfo{year}{2007}).

\bibitem[{\citenamefont{{Dodelson}}(2003)}]{Dodelson2003}
\bibinfo{author}{\bibfnamefont{S.}~\bibnamefont{{Dodelson}}},
  \emph{\bibinfo{title}{{Modern cosmology}}} (\bibinfo{publisher}{Amsterdam
  Netherlands, Academic Press}, \bibinfo{year}{2003}).

\bibitem[{\citenamefont{{Hamilton}}(2000)}]{Hamilton2000}
\bibinfo{author}{\bibfnamefont{A.~J.~S.} \bibnamefont{{Hamilton}}},
  \bibinfo{journal}{\mnras} \textbf{\bibinfo{volume}{312}},
  \bibinfo{pages}{257} (\bibinfo{year}{2000}).

\bibitem[{\citenamefont{{Tegmark} et~al.}(2002)\citenamefont{{Tegmark},
  {Dodelson}, {Eisenstein}, {Narayanan}, {Scoccimarro}, {Scranton}, {Strauss},
  {Connolly}, {Frieman}, {Gunn} et~al.}}]{Tegmark2002}
\bibinfo{author}{\bibfnamefont{M.}~\bibnamefont{{Tegmark}}},
  \bibinfo{author}{\bibfnamefont{S.}~\bibnamefont{{Dodelson}}},
  \bibinfo{author}{\bibfnamefont{D.~J.} \bibnamefont{{Eisenstein}}},
  \bibinfo{author}{\bibfnamefont{V.}~\bibnamefont{{Narayanan}}},
  \bibinfo{author}{\bibfnamefont{R.}~\bibnamefont{{Scoccimarro}}},
  \bibinfo{author}{\bibfnamefont{R.}~\bibnamefont{{Scranton}}},
  \bibinfo{author}{\bibfnamefont{M.~A.} \bibnamefont{{Strauss}}},
  \bibinfo{author}{\bibfnamefont{A.}~\bibnamefont{{Connolly}}},
  \bibinfo{author}{\bibfnamefont{J.~A.} \bibnamefont{{Frieman}}},
  \bibinfo{author}{\bibfnamefont{J.~E.} \bibnamefont{{Gunn}}},
  \bibnamefont{et~al.}, \bibinfo{journal}{\apj} \textbf{\bibinfo{volume}{571}},
  \bibinfo{pages}{191} (\bibinfo{year}{2002}).

\bibitem[{\citenamefont{{Baleisis}
  et~al.}(1998{\natexlab{b}})\citenamefont{{Baleisis}, {Lahav}, {Loan}, and
  {Wall}}}]{Baleisis1998}
\bibinfo{author}{\bibfnamefont{A.}~\bibnamefont{{Baleisis}}},
  \bibinfo{author}{\bibfnamefont{O.}~\bibnamefont{{Lahav}}},
  \bibinfo{author}{\bibfnamefont{A.~J.} \bibnamefont{{Loan}}},
  \bibnamefont{and} \bibinfo{author}{\bibfnamefont{J.~V.}
  \bibnamefont{{Wall}}}, \bibinfo{journal}{\mnras}
  \textbf{\bibinfo{volume}{297}}, \bibinfo{pages}{545}
  (\bibinfo{year}{1998}{\natexlab{b}}).

\bibitem[{\citenamefont{{Peebles}}(1973)}]{Peebles1973}
\bibinfo{author}{\bibfnamefont{P.~J.~E.} \bibnamefont{{Peebles}}},
  \bibinfo{journal}{\apj} \textbf{\bibinfo{volume}{185}}, \bibinfo{pages}{413}
  (\bibinfo{year}{1973}).

\bibitem[{\citenamefont{{Adelman-McCarthy} and the
  SDSS~Collaboration}(2007)}]{Adelman2007}
\bibinfo{author}{\bibfnamefont{J.~K.} \bibnamefont{{Adelman-McCarthy}}}
  \bibnamefont{and} \bibinfo{author}{\bibnamefont{the SDSS~Collaboration}},
  \bibinfo{journal}{ArXiv e-prints}  (\bibinfo{year}{2007}),
  \eprint{0707.3413}.

\bibitem[{\citenamefont{{Fukugita} et~al.}(1996)\citenamefont{{Fukugita},
  {Ichikawa}, {Gunn}, {Doi}, {Shimasaku}, and {Schneider}}}]{Fukugita1996}
\bibinfo{author}{\bibfnamefont{M.}~\bibnamefont{{Fukugita}}},
  \bibinfo{author}{\bibfnamefont{T.}~\bibnamefont{{Ichikawa}}},
  \bibinfo{author}{\bibfnamefont{J.~E.} \bibnamefont{{Gunn}}},
  \bibinfo{author}{\bibfnamefont{M.}~\bibnamefont{{Doi}}},
  \bibinfo{author}{\bibfnamefont{K.}~\bibnamefont{{Shimasaku}}},
  \bibnamefont{and} \bibinfo{author}{\bibfnamefont{D.~P.}
  \bibnamefont{{Schneider}}}, \bibinfo{journal}{\aj}
  \textbf{\bibinfo{volume}{111}}, \bibinfo{pages}{1748} (\bibinfo{year}{1996}).

\bibitem[{\citenamefont{{Gunn} et~al.}(1998)\citenamefont{{Gunn}, {Carr},
  {Rockosi}, {Sekiguchi}, {Berry}, {Elms}, {de Haas}, {Ivezi{\'c}}, {Knapp},
  {Lupton} et~al.}}]{Gunn1998}
\bibinfo{author}{\bibfnamefont{J.~E.} \bibnamefont{{Gunn}}},
  \bibinfo{author}{\bibfnamefont{M.}~\bibnamefont{{Carr}}},
  \bibinfo{author}{\bibfnamefont{C.}~\bibnamefont{{Rockosi}}},
  \bibinfo{author}{\bibfnamefont{M.}~\bibnamefont{{Sekiguchi}}},
  \bibinfo{author}{\bibfnamefont{K.}~\bibnamefont{{Berry}}},
  \bibinfo{author}{\bibfnamefont{B.}~\bibnamefont{{Elms}}},
  \bibinfo{author}{\bibfnamefont{E.}~\bibnamefont{{de Haas}}},
  \bibinfo{author}{\bibfnamefont{{\v Z}.}~\bibnamefont{{Ivezi{\'c}}}},
  \bibinfo{author}{\bibfnamefont{G.}~\bibnamefont{{Knapp}}},
  \bibinfo{author}{\bibfnamefont{R.}~\bibnamefont{{Lupton}}},
  \bibnamefont{et~al.}, \bibinfo{journal}{\aj} \textbf{\bibinfo{volume}{116}},
  \bibinfo{pages}{3040} (\bibinfo{year}{1998}).

\bibitem[{\citenamefont{{Yahata} et~al.}(2007)\citenamefont{{Yahata},
  {Yonehara}, {Suto}, {Turner}, {Broadhurst}, and {Finkbeiner}}}]{Yahata2007}
\bibinfo{author}{\bibfnamefont{K.}~\bibnamefont{{Yahata}}},
  \bibinfo{author}{\bibfnamefont{A.}~\bibnamefont{{Yonehara}}},
  \bibinfo{author}{\bibfnamefont{Y.}~\bibnamefont{{Suto}}},
  \bibinfo{author}{\bibfnamefont{E.~L.} \bibnamefont{{Turner}}},
  \bibinfo{author}{\bibfnamefont{T.}~\bibnamefont{{Broadhurst}}},
  \bibnamefont{and} \bibinfo{author}{\bibfnamefont{D.~P.}
  \bibnamefont{{Finkbeiner}}}, \bibinfo{journal}{Publications of the
  Astronomical Society of Japan} \textbf{\bibinfo{volume}{59}},
  \bibinfo{pages}{205} (\bibinfo{year}{2007}).

\bibitem[{\citenamefont{{Schlegel} et~al.}(1998)\citenamefont{{Schlegel},
  {Finkbeiner}, and {Davis}}}]{Schlegel1998}
\bibinfo{author}{\bibfnamefont{D.~J.} \bibnamefont{{Schlegel}}},
  \bibinfo{author}{\bibfnamefont{D.~P.} \bibnamefont{{Finkbeiner}}},
  \bibnamefont{and} \bibinfo{author}{\bibfnamefont{M.}~\bibnamefont{{Davis}}},
  \bibinfo{journal}{\apj} \textbf{\bibinfo{volume}{500}}, \bibinfo{pages}{525}
  (\bibinfo{year}{1998}).

\bibitem[{\citenamefont{{Gott} et~al.}(2005)\citenamefont{{Gott}, {Juri{\'c}},
  {Schlegel}, {Hoyle}, {Vogeley}, {Tegmark}, {Bahcall}, and
  {Brinkmann}}}]{Gott2005}
\bibinfo{author}{\bibfnamefont{J.~R.~I.} \bibnamefont{{Gott}}},
  \bibinfo{author}{\bibfnamefont{M.}~\bibnamefont{{Juri{\'c}}}},
  \bibinfo{author}{\bibfnamefont{D.}~\bibnamefont{{Schlegel}}},
  \bibinfo{author}{\bibfnamefont{F.}~\bibnamefont{{Hoyle}}},
  \bibinfo{author}{\bibfnamefont{M.}~\bibnamefont{{Vogeley}}},
  \bibinfo{author}{\bibfnamefont{M.}~\bibnamefont{{Tegmark}}},
  \bibinfo{author}{\bibfnamefont{N.}~\bibnamefont{{Bahcall}}},
  \bibnamefont{and}
  \bibinfo{author}{\bibfnamefont{J.}~\bibnamefont{{Brinkmann}}},
  \bibinfo{journal}{\apj} \textbf{\bibinfo{volume}{624}}, \bibinfo{pages}{463}
  (\bibinfo{year}{2005}).

\bibitem[{\citenamefont{{Eisenstein} and {Hu}}(1999)}]{Eisenstein1999}
\bibinfo{author}{\bibfnamefont{D.~J.} \bibnamefont{{Eisenstein}}}
  \bibnamefont{and} \bibinfo{author}{\bibfnamefont{W.}~\bibnamefont{{Hu}}},
  \bibinfo{journal}{\apj} \textbf{\bibinfo{volume}{511}}, \bibinfo{pages}{5}
  (\bibinfo{year}{1999}).

\bibitem[{\citenamefont{{Stubbs} et~al.}(2004)\citenamefont{{Stubbs},
  {Sweeney}, {Tyson}, and {LSST}}}]{LSST}
\bibinfo{author}{\bibfnamefont{C.~W.} \bibnamefont{{Stubbs}}},
  \bibinfo{author}{\bibfnamefont{D.}~\bibnamefont{{Sweeney}}},
  \bibinfo{author}{\bibfnamefont{J.~A.} \bibnamefont{{Tyson}}},
  \bibnamefont{and} \bibinfo{author}{\bibnamefont{{LSST}}}, in
  \emph{\bibinfo{booktitle}{Bulletin of the American Astronomical Society}}
  (\bibinfo{publisher}{New York, American Institute of Physics},
  \bibinfo{year}{2004}), vol.~\bibinfo{volume}{36}, p. \bibinfo{pages}{1527}.

\bibitem[{\citenamefont{{Ivezic} et~al.}(2006)\citenamefont{{Ivezic}, {Tyson},
  {Strauss}, {Kahn}, {Stubbs}, {Pinto}, {Cook}, and {LSST
  Collaboration}}}]{Ivezic2006}
\bibinfo{author}{\bibfnamefont{Z.}~\bibnamefont{{Ivezic}}},
  \bibinfo{author}{\bibfnamefont{A.~J.} \bibnamefont{{Tyson}}},
  \bibinfo{author}{\bibfnamefont{M.~A.} \bibnamefont{{Strauss}}},
  \bibinfo{author}{\bibfnamefont{S.}~\bibnamefont{{Kahn}}},
  \bibinfo{author}{\bibfnamefont{C.}~\bibnamefont{{Stubbs}}},
  \bibinfo{author}{\bibfnamefont{P.}~\bibnamefont{{Pinto}}},
  \bibinfo{author}{\bibfnamefont{K.}~\bibnamefont{{Cook}}}, \bibnamefont{and}
  \bibinfo{author}{\bibnamefont{{LSST Collaboration}}}, in
  \emph{\bibinfo{booktitle}{Bulletin of the American Astronomical Society}}
  (\bibinfo{publisher}{New York, American Institute of Physics},
  \bibinfo{year}{2006}), vol.~\bibinfo{volume}{38}, p. \bibinfo{pages}{1017}.

\bibitem[{\citenamefont{{Huterer} et~al.}(2006)\citenamefont{{Huterer},
  {Takada}, {Bernstein}, and {Jain}}}]{Huterer2006}
\bibinfo{author}{\bibfnamefont{D.}~\bibnamefont{{Huterer}}},
  \bibinfo{author}{\bibfnamefont{M.}~\bibnamefont{{Takada}}},
  \bibinfo{author}{\bibfnamefont{G.}~\bibnamefont{{Bernstein}}},
  \bibnamefont{and} \bibinfo{author}{\bibfnamefont{B.}~\bibnamefont{{Jain}}},
  \bibinfo{journal}{\mnras} \textbf{\bibinfo{volume}{366}},
  \bibinfo{pages}{101} (\bibinfo{year}{2006}).

\bibitem[{\citenamefont{{Crawford}}(2009)}]{Crawford09}
\bibinfo{author}{\bibfnamefont{F.}~\bibnamefont{{Crawford}}},
  \bibinfo{journal}{\apj} \textbf{\bibinfo{volume}{692}}, \bibinfo{pages}{887}
  (\bibinfo{year}{2009}).

\bibitem[{\citenamefont{{Kaiser} et~al.}(2002)\citenamefont{{Kaiser}, {Aussel},
  {Burke}, {Boesgaard}, {Chambers}, {Chun}, {Heasley}, {Hodapp}, {Hunt},
  {Jedicke} et~al.}}]{PanStarrs}
\bibinfo{author}{\bibfnamefont{N.}~\bibnamefont{{Kaiser}}},
  \bibinfo{author}{\bibfnamefont{H.}~\bibnamefont{{Aussel}}},
  \bibinfo{author}{\bibfnamefont{B.~E.} \bibnamefont{{Burke}}},
  \bibinfo{author}{\bibfnamefont{H.}~\bibnamefont{{Boesgaard}}},
  \bibinfo{author}{\bibfnamefont{K.}~\bibnamefont{{Chambers}}},
  \bibinfo{author}{\bibfnamefont{M.~R.} \bibnamefont{{Chun}}},
  \bibinfo{author}{\bibfnamefont{J.~N.} \bibnamefont{{Heasley}}},
  \bibinfo{author}{\bibfnamefont{K.}~\bibnamefont{{Hodapp}}},
  \bibinfo{author}{\bibfnamefont{B.}~\bibnamefont{{Hunt}}},
  \bibinfo{author}{\bibfnamefont{R.}~\bibnamefont{{Jedicke}}},
  \bibnamefont{et~al.}, in \emph{\bibinfo{booktitle}{Society of Photo-Optical
  Instrumentation Engineers (SPIE) Conference Series}}, edited by
  \bibinfo{editor}{\bibnamefont{{J.~A.~Tyson \& S.~Wolff}}}
  (\bibinfo{publisher}{Bellingham, WA, SPIE}, \bibinfo{year}{2002}), vol.
  \bibinfo{volume}{4836}, p. \bibinfo{pages}{154}.

\bibitem[{\citenamefont{{Kashlinsky} et~al.}(2008)\citenamefont{{Kashlinsky},
  {Atrio-Barandela}, {Kocevski}, and {Ebeling}}}]{Kashlinsky2008ApJL}
\bibinfo{author}{\bibfnamefont{A.}~\bibnamefont{{Kashlinsky}}},
  \bibinfo{author}{\bibfnamefont{F.}~\bibnamefont{{Atrio-Barandela}}},
  \bibinfo{author}{\bibfnamefont{D.}~\bibnamefont{{Kocevski}}},
  \bibnamefont{and}
  \bibinfo{author}{\bibfnamefont{H.}~\bibnamefont{{Ebeling}}},
  \bibinfo{journal}{\apjl} \textbf{\bibinfo{volume}{686}}, \bibinfo{pages}{L49}
  (\bibinfo{year}{2008}).

\bibitem[{\citenamefont{{Chluba} et~al.}(2005)\citenamefont{{Chluba},
  {H{\"u}tsi}, and {Sunyaev}}}]{Chluba2005}
\bibinfo{author}{\bibfnamefont{J.}~\bibnamefont{{Chluba}}},
  \bibinfo{author}{\bibfnamefont{G.}~\bibnamefont{{H{\"u}tsi}}},
  \bibnamefont{and} \bibinfo{author}{\bibfnamefont{R.~A.}
  \bibnamefont{{Sunyaev}}}, \bibinfo{journal}{\aap}
  \textbf{\bibinfo{volume}{434}}, \bibinfo{pages}{811} (\bibinfo{year}{2005}).

\bibitem[{\citenamefont{{Ho} et~al.}(2009)\citenamefont{{Ho}, {Dedeo}, and
  {Spergel}}}]{2009arXiv0903.2845H}
\bibinfo{author}{\bibfnamefont{S.}~\bibnamefont{{Ho}}},
  \bibinfo{author}{\bibfnamefont{S.}~\bibnamefont{{Dedeo}}}, \bibnamefont{and}
  \bibinfo{author}{\bibfnamefont{D.}~\bibnamefont{{Spergel}}},
  \bibinfo{journal}{ArXiv e-prints}  (\bibinfo{year}{2009}),
  \eprint{0903.2845}.

\bibitem[{\citenamefont{{Kashlinsky} et~al.}(2009)\citenamefont{{Kashlinsky},
  {Atrio-Barandela}, {Kocevski}, and {Ebeling}}}]{2009ApJ...691.1479K}
\bibinfo{author}{\bibfnamefont{A.}~\bibnamefont{{Kashlinsky}}},
  \bibinfo{author}{\bibfnamefont{F.}~\bibnamefont{{Atrio-Barandela}}},
  \bibinfo{author}{\bibfnamefont{D.}~\bibnamefont{{Kocevski}}},
  \bibnamefont{and}
  \bibinfo{author}{\bibfnamefont{H.}~\bibnamefont{{Ebeling}}},
  \bibinfo{journal}{\apj} \textbf{\bibinfo{volume}{691}}, \bibinfo{pages}{1479}
  (\bibinfo{year}{2009}).

\bibitem[{\citenamefont{Kashlinsky et~al.}(2010)\citenamefont{Kashlinsky,
  Atrio-Barandela, Ebeling, Edge, and Kocevski}}]{Kashlinsky09}
\bibinfo{author}{\bibfnamefont{A.}~\bibnamefont{Kashlinsky}},
  \bibinfo{author}{\bibfnamefont{F.}~\bibnamefont{Atrio-Barandela}},
  \bibinfo{author}{\bibfnamefont{H.}~\bibnamefont{Ebeling}},
  \bibinfo{author}{\bibfnamefont{A.}~\bibnamefont{Edge}}, \bibnamefont{and}
  \bibinfo{author}{\bibfnamefont{D.}~\bibnamefont{Kocevski}},
  \bibinfo{journal}{Astrophys. J.} \textbf{\bibinfo{volume}{712}},
  \bibinfo{pages}{L81} (\bibinfo{year}{2010}).

\bibitem[{\citenamefont{{Ostriker} and {Vishniac}}(1986)}]{1986ApJ...306L..51O}
\bibinfo{author}{\bibfnamefont{J.~P.} \bibnamefont{{Ostriker}}}
  \bibnamefont{and} \bibinfo{author}{\bibfnamefont{E.~T.}
  \bibnamefont{{Vishniac}}}, \bibinfo{journal}{\apjl}
  \textbf{\bibinfo{volume}{306}}, \bibinfo{pages}{L51} (\bibinfo{year}{1986}).

\bibitem[{\citenamefont{{Scharf} et~al.}(1992)\citenamefont{{Scharf},
  {Hoffman}, {Lahav}, and {Lynden-Bell}}}]{Scharf1992}
\bibinfo{author}{\bibfnamefont{C.}~\bibnamefont{{Scharf}}},
  \bibinfo{author}{\bibfnamefont{Y.}~\bibnamefont{{Hoffman}}},
  \bibinfo{author}{\bibfnamefont{O.}~\bibnamefont{{Lahav}}}, \bibnamefont{and}
  \bibinfo{author}{\bibfnamefont{D.}~\bibnamefont{{Lynden-Bell}}},
  \bibinfo{journal}{\mnras} \textbf{\bibinfo{volume}{256}},
  \bibinfo{pages}{229} (\bibinfo{year}{1992}).

\bibitem[{\citenamefont{Jackson}(1999)}]{Jackson}
\bibinfo{author}{\bibfnamefont{J.~D.} \bibnamefont{Jackson}},
  \emph{\bibinfo{title}{Classical electrodynamics; 3rd ed.}}
  (\bibinfo{publisher}{Wiley}, \bibinfo{address}{New York, NY},
  \bibinfo{year}{1999}).

\bibitem[{\citenamefont{Copi et~al.}(2004)\citenamefont{Copi, Huterer, and
  Starkman}}]{Copi2004}
\bibinfo{author}{\bibfnamefont{C.~J.} \bibnamefont{Copi}},
  \bibinfo{author}{\bibfnamefont{D.}~\bibnamefont{Huterer}}, \bibnamefont{and}
  \bibinfo{author}{\bibfnamefont{G.~D.} \bibnamefont{Starkman}},
  \bibinfo{journal}{Phys. Rev. D} \textbf{\bibinfo{volume}{70}},
  \bibinfo{pages}{043515} (\bibinfo{year}{2004}).

\end{thebibliography}

\end{document}